\documentclass[reprint,amsmath,amssymb,aps,prd]{revtex4-2}

\usepackage{dcolumn}
\usepackage{subcaption}
\usepackage{hyperref}
\usepackage[pdftex]{graphicx}
\usepackage{amsmath, amssymb, amsthm, braket,mathrsfs}
\usepackage[T1]{fontenc}
\usepackage{orcidlink}
\usepackage{bm}
\hypersetup{colorlinks=true}
\usepackage[normalem]{ulem}

\DeclareMathOperator{\Tr}{Tr}

\begin{document}


\preprint{APS/123-QED}

\title{Topological transition from a hopfion to a toron via flexoelectric self-polarization in chiral liquid crystals}
\author{Paul Leask\orcidlink{0000-0002-6012-0034}}\email{palea@kth.se}
\affiliation{Department of Physics, KTH Royal Institute of Technology, 10691 Stockholm, Sweden}

\date{\today}

\begin{abstract}
The presence of topological defects in apolar chiral liquid crystals cause orientational distortions, leading to non-uniform strain.
This non-uniform strain generates an electric polarization response due to the flexoelectric effect, which induces an internal electric field.
Associated to this electric field is an electrostatic self-energy, which has a back-reaction on the director field.
Calculation of this internal electric field and its resulting back-reaction on the director field is complicated.
We propose a method to do such, adapting a method recently developed to study the magnetostatic self-interaction effect on skyrmions in chiral ferromagnets.
Bloch skyrmions in chiral magnets are solenoidal and are unaffected by the magnetostatic self-interaction.
However, Bloch skyrmions in liquid crystals yield non-solenoidal flexoelectric polarization and, thus, are affected by the electrostatic self-interaction.
Additionally, as the flexoelectric coefficients are increased in strength, a transition from a hopfion to a toron is observed in three-dimensional confined systems.
\end{abstract}

\maketitle


\section{Introduction}
\label{sec: Introduction}

Liquid crystals are unique states of soft matter that exhibit properties between those of conventional liquids and solid crystals.
They are characterized by the long-range order of their molecular orientations, described by the director field, allowing for fascinating physical phenomena and applications.
The director field encodes spatial variations of the local average molecular alignment direction of the constituent chiral molecules.
By anchoring the liquid crystal at solid boundaries, the director can easily be distorted, giving rise to a myriad of topological defects in the liquid crystal \cite{Smalyukh_2023}.
This homeotropic anchoring introduces geometric frustration to the system \cite{Selinger_2017}, where the boundary conditions are incompatible with the favored cholesteric twist.
It is because of this geometric frustration that topological defects form.
Among the multitude of topological defects that can emerge in liquid crystals, skyrmions and hopfions have gained significant interest recently.

Skyrmions were first studied in condensed matter systems, in the context of chiral magnets \cite{Bogdanov_1989,Bogdanov_1994}.
In those systems, skyrmions are topological spin defects in the magnetization and behave like magnetic quasi-particles.
They were later predicted to exist in chiral liquid crystals \cite{Bogdanov_1998} and nematic liquid crystals \cite{Bogdanov_2003}, where the skyrmions in these systems are defects in the director field that also exhibit particle-like properties.
In addition to skyrmions, the emergence of hopfions in condensed matter systems has gathered great interest \cite{Tai_2018,Tai_2020,Kuchkin_2023,Leonov_2023}.
Unlike a skyrmion, which can be viewed as a localized vortex-like state with a defined core \cite{Bogdanov_1989}, a hopfion can be interpreted as a twisted skyrmion string, forming a closed loop in real space \cite{Sutcliffe_2018}.
They have also been realized experimentally in liquid crystal systems \cite{Ackerman_2015,Ackerman_2017}.

While skyrmions in chiral magnets and chiral liquid crystals are similar in nature, they have distinct manifestations and mechanisms of formation.
In chiral magnets, skyrmions are formed due to the competition between exchange interactions and Dzyaloshinskii--Moriya interactions, which arise from the lack of inversion symmetry in the crystal lattice.
So, skyrmions in chiral magnets are driven primarily by magnetic interactions, whereas those in liquid crystals arise from molecular orientational order.
Essentially, the difference between the two physical systems is spin orientations versus molecular orientations.

Both skyrmions in chiral magnets and in chiral liquid crystals share a common topological nature characterized by a topological homotopy invariant, ensuring stability against perturbations.
While their manifestation is unique in both systems, the fundamental stabilization mechanism responsible for skyrmion formation is the same in both systems.
This relationship between skyrmions in chiral magnets and chiral liquid crystals was shown by Leonov \textit{et al}. \cite{Leonov_2014}.

In magnetic systems, the magnetization induces an internal magnetic field known as the stray, or \emph{demagnetizing}, field \cite{Hubert_2009}, which acts to reduce the total magnetic moment.
Associated to this internal demagnetizing field is a magnetostatic self-energy, generated by dipolar self-interactions of the magnetization.
Determining the stray field and its back-reaction on the magnetization is a difficult task \cite{Fratta_2020} and it can lead to the stabilization/destabilization of topological defect structures \cite{Leask_Speight_2025}.

We study the analogue problem, that is, depolarization in liquid crystals.
The emergence of topological defects in liquid crystals results in orientational distortions of the director field, creating non-uniform strain in the system.
In response to this, an electric polarization is induced, generating an internal electric depolarizing field, similar to the piezoelectric effect \cite{Meyer_1969}.
Akin to the stray field energy, there is an electrostatic self-energy corresponding to the depolarizing field \cite{Meyer_1987}.
Including this electrostatic self-interaction makes the problem non-local.
The author recently developed a method to include the demagnetization field and compute its back-reaction in chiral ferromagnets \cite{Leask_Speight_2025}.
Such methods were also developed recently in the context of knotted string solitons in an extended version of the standard model of particle physics \cite{Hamada_2025}, and the back-reaction of the Coulomb force in an extension of the Skyrme model \cite{Gudnason_2025}.
Our aim is to adapt these methods developed originally in models of particle physics and condensed matter systems to soft matter systems.


\section{The chiral liquid crystal model}
\label{sec: The model}

The system that we wish to model is an apolar chiral liquid crystal, described by a director field $\mathbf{n}(\mathbf{x})\in \mathbb{R}P^2 \cong S^2/\mathbb{Z}_2$ \cite{DeGennes_1993,Stephen_1974}.
That is, the director $\mathbf{n}$ is a vector in $\mathbb{R}^3$ of unit length $|\mathbf{n}|=1$, such that $\mathbf{n}$ and $-\mathbf{n}$ describe the same state, since the director $\mathbf{n}$ is the average molecular alignment direction.
Let us define the thin disk $\Omega_R^d$ of radius $R$ and thickness $d$ by
\begin{equation}
    \Omega_R^d = \left\{(x,y,z) \in\mathbb{R}^3 \,:\, |(x,y)| < R, \, |z|\leq \frac{d}{2}\right\}.
\end{equation}
We will restrict our system to the confined geometry
\begin{equation}
\label{eq: Confined geometry}
    \Omega \equiv \lim_{R\rightarrow\infty} \Omega^d_R = \left\{ (x,y,z) \in\mathbb{R}^3 \,:\, |z|\leq \frac{d}{2}  \right\},
\end{equation}
where $d$ is the cell gap under an external electric field with homeotropic anchoring \cite{DeMatteis_2019}.
Let us first begin by considering a non-chiral, apolar nematic liquid crystal.
In the simplest approximation with isotropic elasticity, the Frank--Oseen energy for such a liquid crystal takes the simple harmonic form
\begin{equation}
\label{eq: Basic FO energy}
    F_{\textup{FO}} = \frac{1}{2} K\int_{\mathbb{R}^3} \textup{d}^3x |\nabla\mathbf{n}|^2.
\end{equation}
Director configurations which minimize this energy are solutions of the Laplace equation $\Delta\mathbf{n}=\mathbf{0}$.
Metastable inhomogeneous solutions of this were found by Belavin and Polyakov in the context of isotropic ferromagnets \cite{Belavin-Polyakov}.

More insight can be gained by considering the elastic modes.
We can decompose the director gradient tensor $\nabla\mathbf{n}$ into four normal modes $(\mathbf{B},T,S,\mathbf{\Delta})$, these are distinct irreducible representations (irreps) of the rotation group.
The director gradient tensor can then be expressed in terms of these normal modes as \cite{Selinger_2018}
\begin{equation}
\label{eq: Normal mode decomposition}
    \partial_i n_j = -n_i B_j + \frac{1}{2}T\epsilon_{ijk}n_k + \frac{1}{2}S(\delta_{ij}-n_in_j) + \Delta_{ij}.
\end{equation}
In liquid crystal physics terminology, the standard bend vector is
\begin{equation}
    \mathbf{B} = -(\mathbf{n}\cdot\bm{\nabla})\mathbf{n} = \mathbf{n} \times (\bm{\nabla}\times\mathbf{n}),
\end{equation}
the standard pseudoscalar twist is
\begin{equation}
    T = \mathbf{n} \cdot (\bm{\nabla}\times\mathbf{n}),
\end{equation}
and the standard splay vector is \cite{Lavrentovich_2018}
\begin{equation}
    \mathbf{S} = S\mathbf{n}, \quad S = \bm{\nabla}\cdot\mathbf{n}.
\end{equation}
The last normal mode is the traceless symmetric tensor $\mathbf{\Delta}$, known as the biaxial splay \cite{Selinger_2018}.
Note that the biaxial splay $\mathbf{\Delta}$ is not to be confused with the Laplace operator $\Delta=-\nabla^2$.
It is a symmetric traceless tensor such that $\mathbf{\Delta}\mathbf{n}=\mathbf{0}$.
Let $q$ be the positive eigenvalue of $\mathbf{\Delta}$ and $\mathbf{n}_1$, $\mathbf{n}_2$ be the eigenvectors orthogonal to $\mathbf{n}$.
Then the biaxial splay can be given by \cite{DeMatteis_2020}
\begin{equation}
    \mathbf{\Delta} = q (\mathbf{n}_1 \otimes \mathbf{n}_1 - \mathbf{n}_2 \otimes \mathbf{n}_2).
\end{equation}
Inserting the normal mode decomposition \eqref{eq: Normal mode decomposition} into the Frank--Oseen energy \eqref{eq: Basic FO energy} yields
\begin{equation}
    F_{\textup{FO}} = \frac{1}{2} K\int_{\mathbb{R}^3} \textup{d}^3x \left\{ \frac{1}{2}|\mathbf{S}|^2 + \frac{1}{2}T^2 + |\mathbf{B}|^2 + \Tr(\mathbf{\Delta}^2) \right\}.
\end{equation}
It is clear that all four modes cost elastic free energy.
In particular, the energy cost of splay, twist and bend deformations are equivalent.

Let us now consider a liquid crystal composed of chiral molecules, with different elastic deformation costs, in the confined geometry \eqref{eq: Confined geometry}.
Then the chirality of these molecules is characterized by some pseudoscalar $q_0$ that couples to the twist $T$.
The associated Frank--Oseen free energy, neglecting the energy cost due to saddle-splay, can be expressed as \cite{Chaikin_Lubensky_1995,Ackerman_2014}
\begin{align}
    F_{\textup{FO}} = \, & \int_{\Omega} \textup{d}^3x \left\{ \frac{1}{2}K_1 |\mathbf{S}|^2 + \frac{1}{2}K_2 (T+q_0)^2 + \frac{1}{2}K_3 |\mathbf{B}|^2 \right\} \nonumber \\
    = \, & \int_{\Omega} \textup{d}^3x \left\{ \frac{K_1}{2} (\bm{\nabla}\cdot \mathbf{n})^2 + \frac{K_2}{2} \left[ \mathbf{n} \cdot (\bm{\nabla}\times\mathbf{n}) + \frac{2\pi}{p} \right]^2 \right. \nonumber \\
    & \left. + \frac{K_3}{2} \left[\mathbf{n} \times (\bm{\nabla} \times \mathbf{n})\right]^2\right\},
\end{align}
where $q_0=2\pi/p$ is the cholesteric twist and $p$ is the cholesteric pitch with a defined length at which the director twists by $2\pi$.
The cholesteric phase is characterized by the presence of enantiomorphy ($q_0\neq0$), and is distinguishable from the nematic phase ($q_0=0$).
The Frank elastic constants $K_1$, $K_2$, and $K_3$ determine the energy cost of splay, twist, and bend deformations, respectively.
Skyrmion solutions in nematic liquid crystal ($q_0=0$) were first theorized in \cite{Bogdanov_2003}.

Chiral liquid crystals are dielectric materials that respond to external electric fields.
This generates a corresponding coupled electric energy of the form \cite{Ackerman_2017,Kim_2016}
\begin{equation}
    \mathcal{E}_{\textup{elec}}=-\frac{\epsilon_0 \Delta\epsilon}{2} (\mathbf{E}_{\textup{ext}} \cdot \mathbf{n})^2
\end{equation}
where $\mathbf{E}_{\textup{ext}}$ is the external electric field, $\epsilon_0$ is the vacuum permittivity and $\Delta\epsilon$ is the dielectric anisotropy.
We will only consider the applied electric field orthogonal to topological defects in the $(x,y)$-plane, that is $\mathbf{E}_{\textup{ext}}=(0,0,E_z)$.

In experimental realizations, liquid crystals are placed between parallel plates with a potential difference.
This imposes boundary conditions orthogonal to the plates on the liquid crystal director field.
In particular, this can impose strong homeotropic anchoring conditions \cite{Selinger_2017}
\begin{equation}
\label{eq: Strong homeotropic anchoring}
    \mathbf{n}(x,y,z=\pm d/2)= \mathbf{e}_z = (0,0,1).
\end{equation}
This can be accounted for in two dimensional systems by including the Rapini--Papoular homeotropic surface anchoring potential \cite{Duzgun_2018,Smalyukh_2020}
\begin{equation}
    \mathcal{E}_{\textup{anch}} = -\frac{1}{2}W_0 n_z^2,
\end{equation}
where $W_0$ is the effective surface anchoring strength which favors director alignment in the $z$-direction, that is $\mathbf{n}=\pm\mathbf{e}_{z}=(0,0,\pm1)$ director configurations.
This term acts to mimic homeotropic anchoring conditions at the cell surfaces of a three-dimensional system.

The Frank--Oseen free energy we are interested in, including the electric energy and homeotropic anchoring, is given by the energy functional
\begin{align}
\label{eq: Frank-Oseen energy}
    &F_{\textup{FO}} = \,  \int_{\Omega} \textup{d}^3x \left\{ \frac{K_{1}}{2} (\bm{\nabla}\cdot \mathbf{n})^2 +  \frac{K_{2}}{2} \left[ \mathbf{n} \cdot (\bm{\nabla}\times\mathbf{n}) \right]^2 \right. \nonumber \\
    \, & \left. + \frac{K_{3}}{2} \left[\mathbf{n} \times \bm{\nabla} \times \mathbf{n}\right]^2 + K_2 q_0\left[ \mathbf{n} \cdot (\bm{\nabla}\times\mathbf{n}) \right] + V(\mathbf{n})\right\},
\end{align}
where the potential is $V(\mathbf{n}) = \mathcal{E}_{\textup{elec}}$ in three dimensions and $V(\mathbf{n})=\mathcal{E}_{\textup{elec}}+\mathcal{E}_{\textup{anch}}$ in two dimensions.
While this model does account for an external applied electric field, it does not include the electrostatic self-interaction energy arising from the flexoelectric effect.
We will now show how to do this.


\section{Flexoelectric polarization}
\label{sec: Flexoelectric polarization}

When liquid crystals possess macroscopic electric polarization $\mathbf{P}_f$ (where $\mathbf{P}_f$ is spontaneous or induced by some external, non-electric field related, factors), then they induce a linear-in-field energy contribution \cite{Blinov_2017}.
One such source of macroscopic electric polarization generation is related to orientational distortions in liquid crystals.
The case we consider here is molecules with permanent dipole moments.
This leads to piezoelectric effects and generates a dipolar piezoelectric-like polarization.
However, piezoelectricity is due to uniform strain, whereas this polarization is caused by the mechanical curvature, or flexion, of the director field $\mathbf{n}$, and is called \textit{flexoelectric} \cite{DeGennes_1993}.
If we fix the splay and induce polarization, this generates an additional energy contribution given by
\begin{equation}
    F_{\textup{S}} = \frac{1}{2}K_1 \left| \mathbf{n}(\bm{\nabla}\cdot\mathbf{n}) - c_1\mathbf{P} \right|^2 + \frac{1}{2}\mu|\mathbf{P}|^2.
\end{equation}
Varying this with respect to the polarization yields \cite{Selinger_2025}
\begin{equation}
    \frac{\delta F_{\textup{S}}}{\delta \mathbf{P}} = 0 \quad \Rightarrow \quad \mathbf{P} = \left( \frac{c_1 K_1}{c_1^2 K_1 + \mu} \right) (\bm{\nabla}\cdot\mathbf{n})\mathbf{n}.
\end{equation}
Similarly, we can do the same thing for fixed-bend induced polarization, which generates a polarization
\begin{equation}
    \mathbf{P} = \left( \frac{c_3 K_3}{c_3^2 K_3 + \mu} \right) [ \mathbf{n} \times (\bm{\nabla} \times \mathbf{n})].
\end{equation}
A formal theory of these flexoelectric effects was developed by Meyer \cite{Meyer_1969}.
This can be expressed as \cite{Meyer_1987,Krekhov_2018}
\begin{equation}
\label{eq: Flexoelectric polarization}
    \mathbf{P}_f = e_1 \left[ (\bm{\nabla} \cdot \mathbf{n}) \mathbf{n} \right] + e_3 \left[ \mathbf{n} \times (\bm{\nabla} \times \mathbf{n}) \right],
\end{equation}
where $e_1$ and $e_3$ are, respectively, the piezoelectric constants for the splay and bend of the molecules \cite{Kagawa_1977}.
These coefficients are material dependent and can be measured directly via the electric current produced by the periodic mechanical flexing of the liquid crystals bounding surfaces \cite{Harden_2006}.

Analogous to demagnetization in chiral magnets, the flexoelectric polarization produces internal sources of electric fields \cite{Meyer_1969}, i.e. it induces an electric dipole moment $\mathbf{p}$, where $\mathbf{p}=\mathbf{P}_f$.
In fact, it generates a continuous electric dipole moment distribution $\mathbf{P}_f: \Omega \rightarrow \mathbb{R}^3$, where $\Omega \subseteq \mathbb{R}^3$ is the confined geometry \eqref{eq: Confined geometry}.
The electric potential $\varphi:\Omega\rightarrow\mathbb{R}$ associated to this continuous dipole distribution induces an internal electric field $\mathbf{E} = -\bm{\nabla}\varphi$. 
It satisfies a Poisson equation for electrostatics \cite{Zavvou_2022,Yang_2022}
\begin{equation}
\label{eq: Poisson equation}
    \Delta\varphi = -\nabla^2\varphi = \frac{1}{\epsilon_0} \rho, \quad \rho = -(\bm{\nabla}\cdot\mathbf{P}_f),
\end{equation}
where $\rho_e$ is the electric charge density and the Laplacian is $\Delta = -\nabla^2$ on $\mathbb{R}^3$.

Using the definition of the electric potential, we see that Gauss' law is
\begin{equation}
    \bm{\nabla} \cdot \mathbf{E} = \frac{\rho}{\epsilon_0}, \quad \rho = -(\bm{\nabla}\cdot\mathbf{P}_f)
\end{equation}
where $\rho$ is the electric charge density.
The electric field induced by the dipole distribution $\mathbf{P}_f$ coincides, therefore, with the electric field induced by the charge distribution $-(\nabla\cdot \mathbf{P}_f)$.
Hence, we may think of $-(\nabla\cdot\mathbf{P}_f)$ as an electric charge density.
Furthermore, we can write
\begin{equation}
    \bm{\nabla} \cdot \left( \epsilon_0 \mathbf{E} + \mathbf{P}_f \right) = \bm{\nabla} \cdot\mathbf{D} = 0,
\end{equation}
where $\mathbf{D}$ is the electric displacement field.
Hence, there is no space charge.

Suppose we have a pair of electric dipole moments $\mathbf{P}_f^{(1)}$ and $\mathbf{P}_f^{(2)}$.
Their interaction energy is
\begin{equation}
    E_{\textup{int}} = -\mathbf{P}_f^{(1)}\cdot\mathbf{E}^{(2)} = -\mathbf{P}_f^{(2)}\cdot\mathbf{E}^{(1)},
\end{equation}
where $\mathbf{E}^{(2)}$ is the electric field induced by the polarization $\mathbf{P}_f^{(2)}$ at the position of $\mathbf{P}_f^{(1)}$, and vice versa.
Therefore, the flexoelectric energy coincides with the energy of a continuous dipole density distribution, which is \cite{Morozovska_2018,Škarabot_2022}
\begin{equation}
\label{eq: Electrostatic self-energy}
    F_{\textup{flexo}} = -\frac{1}{2} \int_{\Omega} \textup{d}^3\mathbf{x} \, \mathbf{E}(\mathbf{x}) \cdot \mathbf{P}_f(\mathbf{x}),
\end{equation}
where $\mathbf{E}$ is the induced electric field \eqref{eq: Electric field}.
We will later want to compute the variation of the flexoelectric energy with respect to the director field $\mathbf{n}$.
For this reason, it proves more useful to express the flexoelectric energy in terms of the scalar electric potential $\varphi$,
\begin{align}\label{molasm}
    F_{\textup{flexo}} = \, & \frac{1}{2} \int_\Omega \textup{d}^3\mathbf{x} \, \mathbf{P}_f \cdot \bm{\nabla}\varphi \nonumber \\
    = \, & -\frac{1}{2} \int_\Omega \textup{d}^3\mathbf{x} \, \left( \bm{\nabla} \cdot \mathbf{P}_f \right) \varphi + \frac{1}{2} \oint_{\partial\Omega} \textup{d}\mathbf{s} \cdot \left( \varphi\mathbf{P}_f \right) \nonumber \\
    = \, & \frac{\epsilon_0}{2} \int_{\mathbb{R}^3} \textup{d}^3\mathbf{x} \, \varphi \Delta\varphi + \frac{1}{2} \oint_{\partial\Omega} \textup{d}\mathbf{s} \cdot \left( \varphi\mathbf{P}_f \right),
\end{align}
by the Divergence Theorem.
We will restrict ourselves to situations where the boundary conditions ensure the boundary term vanishes.
In this case, the flexoelectric energy $F_{\textup{flexo}}$ coincides with the electrostatic self-energy of the charge distribution $-(\nabla\cdot\mathbf{P}_f)$.
To see this, we use the general identity $\varphi\Delta\varphi = \bm{\nabla}\varphi\cdot\bm{\nabla}\varphi - \bm{\nabla}\cdot \left( \varphi\bm{\nabla}\varphi \right)$ and the divergence theorem to express the flexoelectric energy as
\begin{align}
    F_{\textup{flexo}} = \, \frac{\epsilon_0}{2}\int_{\mathbb{R}^3} \textup{d}^3\mathbf{x} \, |\bm{\nabla}\varphi|^2 
    = \,  \frac{\epsilon_0}{2}\int_{\mathbb{R}^3} \textup{d}^3\mathbf{x} \, |\mathbf{E}|^2.
\end{align}

We now detail the two cases of interest.
The first is the flexoelectric self-interaction energy of a translation invariant skyrmion in a chiral liquid crystal, and the second is a hopfion in the confined geometry \eqref{eq: Confined geometry}.


\subsection{Translation invariant skyrmion system}
\label{subsec: Translation invariant skyrmion system}

We impose the translation invariance to be along the $z$-direction, such that the director field $\mathbf{n}$ is independent of $z$ and, therefore, so is the polarization $\mathbf{P}_f$.
We assume that there is a energetically preferential orientation for the director $\mathbf{n}=\mathbf{e}_z$ due to some intrinsic anisotropy or an external applied electric field.
Note that we could have equally have chosen $\mathbf{n}=-\mathbf{e}_z$ as the energy is invariant under the transformation $\mathbf{n}\mapsto-\mathbf{n}$.
That is, the directors $\mathbf{n}$ and  $-\mathbf{n}$ describe the same physical state, with energy $F[\mathbf{n}]=F[-\mathbf{n}]$.
We assume that our polarization field has compact support such that there exists $R_0>0$ whereby $\mathbf{n}(x,y)=\mathbf{e}_z$ and, whence, $\mathbf{P}_f(x,y)=\mathbf{0}$ for all $r:=|(x,y)|\geq R_0$.

The total electrostatic self-energy must either vanish or diverge as the polarization field $\mathbf{P}_f$ is translation invariant.
However, if we consider the energy per unit length in the $z$-direction, then this may be finite.
This coincides with the energy of the thick disk $\Omega_R^1=\{\mathbf{x}\: :\: x^2+y^2\leq R^2,\: |z|\leq 1/2\}$ in the limit $R\rightarrow\infty$ \cite{Leask_Speight_2025}.
In this case, the flux of the boundary term $\varphi\mathbf{P}_f$ through the disk wall vanishes since the polarization $\mathbf{P}_f$ vanishes on the boundary.
Furthermore, the flux through the top and bottom of the disk cancel out due to the translation invariance of $\varphi\mathbf{P}_f$. 
Thus, the electrostatic self-energy of the translation invariant system is
\begin{equation}
    F_{\textup{flexo}}=\frac{\epsilon_0}{2}\int_{\mathbb{R}^2}\textup{d}^2x\, \varphi\Delta\varphi,
\end{equation}
where $\Delta=-\partial_1^2-\partial_2^2$ is the usual flat Laplacian on $\mathbb{R}^2$, and the electric scalar potential $\varphi$ satisfies
\begin{equation}\label{monlarsmi}
    \frac{\partial^2 \varphi}{\partial x^2} + \frac{\partial^2 \varphi}{\partial y^2} = \frac{1}{\epsilon_0}\left(\frac{\partial P_1}{\partial x}+\frac{\partial P_2}{\partial y}\right).
\end{equation}

Consider the asmyptotic $r\rightarrow\infty$ behavior of the electric potential $\varphi:\mathbb{R}^2\rightarrow\mathbb{R}$.
Any solution of the Poisson equation $\Delta\varphi=\rho/\epsilon_0$ on the plane has a multipolar expansion
\begin{equation}
    \varphi=-\frac{q}{2\pi\epsilon_0}\log r +O(r^{-1}),
\end{equation}
where $q=\int_{\mathbb{R}^2}\textup{d}^2x\,\rho$ is the total charge.
Such functions are generally logarithmically unbounded.
However, let us define $B_R(\mathbf{y})$ to be the ball of radius $R$ centered at a point $\mathbf{y}$ in $\mathbb{R}^2$, that is
\begin{equation}
    B_R(\mathbf{y}) = \left\{ \mathbf{x}\in\mathbb{R}^2: \, |\mathbf{x}-\mathbf{y}| < R \right\}.
\end{equation}
Then the total charge electric charge in the system must vanish,
\begin{align}
    q = \, & \int_{\mathbb{R}^2}\textup{d}^2x\,\rho = \lim_{R\rightarrow\infty}\oint_{\partial B_R(0)} \textup{d}\mathbf{s} \cdot \mathbf{P}_f = 0,
\end{align}
by the divergence theorem, since the flexoelectric polarization has compact support.
Therefore, we can conclude that $\varphi$ has (at least) a $1/r$ localization. 


\subsection{Hopfion system}
\label{subsec: Hopfion system}

We now consider the three dimensional case with the strong homeotropic anchoring \eqref{eq: Strong homeotropic anchoring}.
This leads to the cell boundary condition for the polarization,
\begin{equation}
    \mathbf{P}_f(r,z=\pm d/2)=e_1(\partial_zn_z)\mathbf{e}_z -e_3\partial_z\mathbf{n}.
\end{equation}
That is, we compute the energy of a thin disk $\Omega^d_R=\{\mathbf{x}\: :\: r^2=x^2+y^2\leq R^2,\: |z|\leq d/2\}$ in the limit $R\rightarrow\infty$.
As in the translation invariant case, the director, and hence the polarization, has compact support in the $(x,y)$-plane for fixed cell height values $z$.
So, the director field tends to a constant on the disk boundary, that is $\mathbf{n}(r\rightarrow\infty,z)=\mathbf{e}_z$.
This means that the self-induced flexoelectric polarization identically vanishes on the disk boundary $\mathbf{P}_f(r\rightarrow\infty,z)=\mathbf{0}$.
Therefore, we must have that the electric scalar potential also vanishes on the disk boundary, $\varphi(r\rightarrow\infty,z)=0$.
In order to compute the electric potential $\varphi:\mathbb{R}^3\rightarrow\mathbb{R}$, we need to extend beyond the cell boundary.
This leads to the interior/exterior interface problem
\begin{align}
    \begin{cases}
        \Delta\varphi = -\frac{1}{\epsilon_0} \bm{\nabla}\cdot\mathbf{P}_f & \textup{in}\,\Omega, \\
        \Delta\varphi = 0 & \textup{in}\,\mathbb{R}^3/\Omega.
    \end{cases}
\end{align}
In three dimensions, we have the multipole expansion
\begin{equation}
    \varphi = - \frac{q}{4\pi\epsilon_0 r} + O(r^{-2}).
\end{equation}
and the regularity condition $\varphi(r,|z|\rightarrow \infty)\rightarrow 0$.


\section{Variation of the flexoelectric energy}

So far, we have shown how to include the electrostatic self-energy and compute the electric scalar potential $\varphi$ by solving Poisson's equation \eqref{eq: Poisson equation} for fixed director field configuration $\mathbf{n}$.
However, we need to compute the back-reaction of the self-induced electric field $\mathbf{E}$ on the director field $\mathbf{n}$.
To do this, we need to calculate the first variation of the flexoelectric energy $F_{\textup{flexo}}(\mathbf{n})$ with respect to the director field $\mathbf{n}$.

Before proceeding with the variation of the flexoelectric energy, we opt to work in dimensionless units.
This will also make numerical simulations more palatable.
Let us consider an energy and length rescaling with $E=E_0\hat{E}$ and $x=L_0\hat{x}$.
We choose to set our length and energy scales as
\begin{equation}
\label{eq: Rescaled energy/length}
    L_0 = \frac{1}{q_0}\frac{K_1}{K_2}, \quad E_0 = \frac{1}{q_0}\frac{K_1^2}{K_2}.
\end{equation}
Then the rescaled energy is 
\begin{align}
    \hat{F}_{\textup{FFO}} = \, & \int_{\Omega} \textup{d}^3x \left\{ \frac{1}{2} (\bm{\nabla}\cdot \mathbf{n})^2 + \frac{1}{2} \frac{K_2}{K_1} \left[ \mathbf{n} \cdot (\bm{\nabla}\times\mathbf{n}) \right]^2 \right. \nonumber \\
        \, & \left. + \frac{1}{2} \frac{K_3}{K_1} (\mathbf{n} \times \bm{\nabla} \times \mathbf{n})^2 + \left[ \mathbf{n} \cdot (\bm{\nabla}\times\mathbf{n}) \right] \right.\nonumber \\
        \, & \left. + \frac{1}{q_0^2} \frac{K_1}{K_2^2} V(\mathbf{n})\right\} + \hat{F}_{\textup{flexo}},
\end{align}
where the rescaled flexoelectric energy is determined to be
\begin{equation}
\label{eq: Rescaled flexoelectric energy}
    \hat{F}_{\textup{flexo}} = \frac{\epsilon}{2} \int_{\Omega} \hat\varphi \Delta_{\hat{x}} \hat\varphi \, \textup{d}^3\hat{x}, \quad \Delta_{\hat{x}} \hat\varphi = -\frac{1}{\epsilon} \bm{\nabla}_{\hat{x}} \cdot \mathbf{P}.
\end{equation}
Here, we have expressed the polarization $\mathbf{P}_f$ in terms of a rescaled polarization $\mathbf{P}$, where
\begin{equation}
    \mathbf{P}_f = \frac{e_1}{L_0} \mathbf{P}, \quad \mathbf{P} =  (\bm{\nabla}_{\hat{x}} \cdot \mathbf{n}) \mathbf{n}  + \frac{e_3}{e_1} [\mathbf{n} \times (\bm{\nabla}_{\hat{x}} \times \mathbf{n})],
\end{equation}
and introduced the dimensionless vacuum electric permittivity
\begin{equation}
    \epsilon = \frac{K_1\epsilon_0}{e_1^2}.
\end{equation}

A necessary requirement for topological solitons to exist in a field theory is the successful evasion of the Hobart--Derrick theorem \cite{Derrick_1964}.
The Hobart--Derrick theorem is a non-existence theorem that states if the energy functional $F[\mathbf{n}]$ is not stationary against spatial rescaling, then $\mathbf{n}$ cannot be a solution of the field equations.
So, let us consider a coordinate rescaling $\mathbf{x} \mapsto \mathbf{x}'=\mu\mathbf{x}$, for some positive scaling parameter $\mu\in\mathbb{R}_{>0}$.
Then the director field necessarily rescales as $\mathbf{n}_\mu=\mathbf{n}(\mu\mathbf{x})$.
We find that the polarization $\mathbf{P}$ transforms as $\mathbf{P}_\mu = \mu \mathbf{P}(\mu\mathbf{x})$ and the necessary electric scalar potential scaling behavior is $\varphi_\mu(\mathbf{x})=\varphi(\mu\mathbf{x})$.

We now focus on the situation at hand: translation invariant configurations in the $\mathbf{e}_z$ direction.
Recall that we are considering the finite energy of a thick slab $\Omega=\mathbb{R}^2\times[0,1]$.
Under the energy and length rescalings \eqref{eq: Rescaled energy/length}, the thick slab gets mapped to $\Omega'=\mathbb{R}^2\times[0,t]$, where $t=L_0^{-1}$ is the rescaled slab thickness.
We then consider the energy per unit length $F/t$ of this system, which is defined by the two-dimensional adimensional energy functional
\begin{align}
\label{eq: Adimensional 2D energy}
    F_{\textup{FFO}} = \, & \int_{\mathbb{R}^2} \textup{d}^2x \left\{ \frac{1}{2} (\bm{\nabla}\cdot \mathbf{n})^2 + \frac{1}{2} \frac{K_2}{K_1} \left[ \mathbf{n} \cdot (\bm{\nabla}\times\mathbf{n}) \right]^2 \right. \nonumber \\
    \, & \left. + \frac{1}{2} \frac{K_3}{K_1} (\mathbf{n} \times \bm{\nabla} \times \mathbf{n})^2 + \left[ \mathbf{n} \cdot (\bm{\nabla}\times\mathbf{n}) \right] \right.\nonumber \\
    \, & \left. + \frac{1}{q_0^2} \frac{K_1}{K_2^2} V(\mathbf{n}) + \frac{\epsilon}{2}\varphi\Delta\varphi \right\},
\end{align}
where the electric potential satisfies the dimensionless Poisson equation
\begin{equation}
\label{eq: Adimensional 2D Poisson equation}
    \Delta \varphi = -\frac{1}{\epsilon} \bm{\nabla} \cdot \mathbf{P}, \quad \mathbf{P} =  (\bm{\nabla} \cdot \mathbf{n}) \mathbf{n}  + \frac{e_3}{e_1} \left[\mathbf{n} \times (\bm{\nabla} \times \mathbf{n})\right].
\end{equation}

Therefore, under the coordinate rescaling $\mathbf{x} \mapsto \mathbf{x}'=\mu\mathbf{x}$, the dimensionless energy functional becomes
\begin{align}
    F_{\textup{FFO}}(\mu) = \, & F_2 + \frac{1}{\mu}F_1 + \frac{1}{\mu^2} F_0 + F_{\textup{flexo}}.
\end{align}
For skyrmions to exist in this model, we require the energy to be stable against spatial rescalings, which yields the Derrick scaling constraint
\begin{equation}
\label{eq: Derrick scaling}
    \left.\frac{\textup{d}F_{\textup{FFO}}}{\textup{d}\mu}\right|_{\mu=1} = -(F_1 + 2F_0) = 0.
\end{equation}
While the potential energy $F_0$ is positive semi-definite, the first order (in spatial derivatives) term $F_1$ can be negative, and thus provide stability.
Hence, the Hobart--Derrick non-existence theorem can be evaded.

We note that the flexoelectric self-energy is scale invariant in two dimensions and is thus unable to provide stability against spatial rescalings.
Whereas, in comparison with chiral ferromagnets, the magnetostatic self-energy there can stabilize skyrmions as it behaves like a potential under coordinate rescalings \cite{Leask_Speight_2025}.

Let $\mathbf{n}_t$ be a smooth variation of $\mathbf{n}=\mathbf{n}_0$ through fields of compact support and define
$\delta\mathbf{n}=\partial_t\mathbf{n}_t|_{t=0}$.
Denote by $\varphi_t$ the associated unique solution of \eqref{eq: Adimensional 2D Poisson equation}
with source $-\frac{1}{\epsilon} \bm{\nabla}\cdot \mathbf{P}_t$ decaying to $0$ at infinity, and $\dot\varphi=\partial_t\varphi_t|_{t=0}$.
It is important to note that, while $\delta\mathbf{n}$ has compact support, neither $\varphi=\varphi_0$ nor $\dot\varphi$ do: as argued above, they are $1/r$ localized.
Following \cite{Leask_Speight_2025}, the variation of $F_{\textup{flexo}}$ induced by $\mathbf{n}_t$ is found to be given by
\begin{align}
    \frac{\textup{d}}{\textup{d}t}\bigg|_{t=0}F_{\textup{flexo}}(\mathbf{n}_t) = \, & \int_{\mathbb{R}^2}\textup{d}^2x\, (\textup{grad}_{\mathbf{n}}\,F_{\textup{flexo}}) \cdot \delta\mathbf{n},
\end{align}
where the corresponding gradient is
\begin{align}
    \textup{grad}_{\mathbf{n}}\,F_{\textup{flexo}} = \, & \frac{e_3}{e_1} \left[  \left( (\bm{\nabla} \times \mathbf{n}) \times \bm{\nabla}\varphi \right) + \left( \bm{\nabla} \times (\bm{\nabla}\varphi\times\mathbf{n}) \right) \right] \nonumber \\
    \, & - \bm{\nabla}(\bm{\nabla}\varphi \cdot \mathbf{n}) + (\bm{\nabla}\cdot\mathbf{n})\bm{\nabla}\varphi.
\end{align}

We can easily extend the analysis in the previous section to the three-dimensional confined geometry \eqref{eq: Confined geometry}.
The variation is found to be
\begin{align}
    \frac{\textup{d}}{\textup{d}t}\bigg|_{t=0}F_{\textup{flexo}}(\mathbf{n}_t) = \, & \int_{\Omega}\textup{d}^3x\, \bm{\nabla}\varphi \cdot \dot{\mathbf{P}} - \oint_{\partial\Omega} \textup{d}\mathbf{s} \cdot (\varphi\dot{\mathbf{P}}) \nonumber \\
    \, & -\frac{\epsilon}{2}\oint_{\partial\Omega} \textup{d}\mathbf{s} \cdot \left( \varphi\bm{\nabla}\dot\varphi - \dot\varphi\bm{\nabla}\varphi \right).
\end{align}
We can treat this as a thick disk in the limit where the radius of the disk tends to infinity.
Then, using the same argument as in the two-dimensional case, the surface contribution term must vanish on the curved part of the disk in the infinite radius limit.
We will only consider systems in which this boundary term vanishes, such as the case of hopfions \eqref{eq: Hopfion ansatz} in the confined geometry \eqref{eq: Confined geometry}.
Therefore, the variation of the flexoelectric energy is determined to be
\begin{equation}
    \frac{\textup{d}}{\textup{d}t}\bigg|_{t=0}F_{\textup{flexo}}(\mathbf{n}_t) = \int_{\Omega}\textup{d}^3x\, (\textup{grad}_{\mathbf{n}}\,F_{\textup{flexo}}) \cdot \delta\mathbf{n}.
\end{equation}


\section{Relation to chiral magnets}
\label{sec: Relation to chiral magnets}

The stability of two-dimensional skyrmions in chiral liquid crystals arises from the same mechanism responsible for the existence of skyrmions in chiral ferromagnetic systems \cite{Leonov_2014}.
This is due to the chiral interactions imposed by the handedness of the system.
Consider the one-constant approximation where the bend, splay and twist constants are all equal ($K_i=K$).
This corresponds to an apolar, chiral liquid crystal \cite{Stephen_1974}.
For such liquid crystals in an applied electric field $\mathbf{E}_{\textup{ext}}=(0,0,E_z)$, the free-energy in the one-constant approximation can be reduced to the following expression
\begin{align}
    F_{\textup{FFO}} = \, & \int_{\mathbb{R}^2} \textup{d}^2x \left\{ \frac{1}{2} (\nabla \mathbf{n})^2 + \left[ \mathbf{n} \cdot (\bm{\nabla}\times\mathbf{n}) \right] + \frac{1}{q_0^2} \frac{1}{K} V(\mathbf{n}) \right.\nonumber \\
    \, & \left. + \frac{\epsilon}{2}\varphi\Delta\varphi \right\},
\end{align}
where we have used the identity \cite{Hubert_2009}
\begin{align}
\label{eq: Gradient identity}
    \left( \nabla\mathbf{n}\right)^2 = \, & \left( \bm{\nabla}\cdot\mathbf{n}\right)^2 + \left[ \mathbf{n} \cdot (\bm{\nabla} \times \mathbf{n})\right]^2 + \left[\mathbf{n} \times (\bm{\nabla} \times \mathbf{n}) \right]^2 \nonumber \\
    \, & + \bm{\nabla} \cdot \left[ (\mathbf{n}\cdot\bm{\nabla})\mathbf{n} - (\bm{\nabla}\cdot \mathbf{n})\mathbf{n} \right]
\end{align}
which holds for any unit vector $\mathbf{n}$.
This is the energy density of a chiral ferromagnet in the absence of an external magnetic field with the Dzyaloshinskii--Moriya interaction (DMI) arising from the Dresselhaus spin-orbit coupling (SOC) \cite{Leonov_2011}.
The dielectric anisotropy energy in liquid crystals plays the same role as uniaxial anisotropy in chiral magnets.


\section{Numerical method}
\label{sec: Numerical method}

Our interests lie in computing the self-induced flexoelectric polarization of topological solitons in the above systems.
For simplicity, we will use the one constant approximation from here on out.
We will now detail our method for obtaining solitons in the three dimensional system, but the method reduces easily to the two-dimensional translation invariant case.
Topological solitons in this model are minimizers of the adimensional flexoelectric Frank--Oseen free energy 
\begin{align}
\label{eq: Flexoelectric Frank--Oseen energy}
    F_{\textup{FFO}} = \, & \int_{\Omega} \textup{d}^3x \left\{ \frac{1}{2} (\nabla \mathbf{n})^2 + \left[ \mathbf{n} \cdot (\bm{\nabla}\times\mathbf{n}) \right] + \frac{1}{q_0^2} \frac{1}{K} V(\mathbf{n}) \right.\nonumber \\
    \, & \left. + \frac{1}{2}\mathbf{P}\cdot\bm{\nabla}\varphi\right\},
\end{align}
where the electric scalar potential $\varphi$ is subject to the constraint
\begin{align}
\label{eq: Electric potential cases}
    \begin{cases}
        \Delta\varphi = -\frac{1}{\epsilon} \bm{\nabla}\cdot\mathbf{P} & \textup{in}\,\Omega, \\
        \Delta\varphi = 0 & \textup{in}\,\mathbb{R}^3/\Omega.
    \end{cases}
\end{align}
The adimensional polarization is
\begin{equation}
\label{eq: Adimensional polarization}
    \mathbf{P} =  (\bm{\nabla} \cdot \mathbf{n}) \mathbf{n}  + \frac{e_3}{e_1} \left[\mathbf{n} \times (\bm{\nabla} \times \mathbf{n})\right].
\end{equation}

We develop a method to find director fields $\mathbf{n}\in\mathbb{R}P^2$ that simultaneously minimize the flexoelectric Frank--Oseen energy \eqref{eq: Flexoelectric Frank--Oseen energy} and solve the electric potential constraint \eqref{eq: Electric potential cases}.
This method is implemented for NVIDIA CUDA architecture and is adapted from a similar non-local method developed to determine skyrmion crystals stabilized by $\omega$-mesons \cite{Leask_Harland_2024,Leask_2024}.

The inclusion of the flexoelectric polarization self-interaction introduces non-locality into the minimization problem.
Now, the minimization method assumes that Poisson's equation is satisfied during each iteration of the algorithm.
We now need to ensure that it indeed is.
That is, during every iteration of the minimization algorithm, the electric potential $\varphi$ must solve Poisson's equation $\Delta\varphi=\frac{1}{\epsilon}\rho$ with the source $\rho=-(\bm{\nabla}\cdot\mathbf{P})$.
This can be approached by reformulating the problem as an unconstrained optimization problem: minimize the functional
\begin{equation} 
    F(\varphi) = \frac{1}{2}\int_{\mathbb{R}^3} \textup{d}^3x\,|\textup{d}\varphi|^2 + \frac{1}{\epsilon} \int_{\Omega} \textup{d}^3x\,\varphi \left( \bm{\nabla}\cdot\mathbf{P} \right)
\end{equation}
with respect to $\varphi$, where the director field $\mathbf{n}$ is fixed and, thus, the divergence of the polarization $\mathbf{P}$ is also fixed, with
\begin{align}
    \bm{\nabla}\cdot\mathbf{P} = \, & \frac{e_3}{e_1} \left\{ (\bm{\nabla}\times\mathbf{n})^2 - \mathbf{n}\cdot[\bm{\nabla}(\bm{\nabla}\cdot\mathbf{n})] + \mathbf{n}\cdot \nabla^2\mathbf{n} \right. \nonumber \\
    \, & \left. + (\bm{\nabla}\cdot\mathbf{n})^2 + \mathbf{n}\cdot[\bm{\nabla}(\bm{\nabla}\cdot\mathbf{n})] \right\}.
\end{align}

We will use a non-linear conjugate gradient method with a line search strategy to solve this unconstrained problem.
The conjugate stepsize is determined using the Fletcher--Reeves method.

To summarize, the full numerical algorithm is implemented as follows:
\begin{enumerate}
    \item Perform a step of the arrested Newton flow method for the director field $\mathbf{n}$ using a 4th order Runge-Kutta method.
    \item Solve Poisson's equation for the electric potential $\varphi$ using nonlinear conjugate gradient descent with the Fletcher--Reeves method.
    \item Compute the total energy of the configuration $(\mathbf{n}_i,\varphi_i)$ and compare to the energy of the previous configuration $(\mathbf{n}_{i-1},\varphi_{i-1})$. If the energy has increased, arrest the flow.
    \item Check the convergence criteria: $\lVert E_{\textup{dis}}(\mathbf{n}) \rVert_\infty<\varepsilon$. If the convergence criteria has been satisfied, then stop the algorithm.
    \item Repeat the process (return to step 1).
\end{enumerate}


\section{Liquid crystal skyrmions}

In our model, liquid crystal skyrmions are smooth maps $\mathbf{n}:\mathbb{R}^2\rightarrow \mathbb{R}P^2$ that minimize the flexoelectric Frank--Oseen energy \eqref{eq: Flexoelectric Frank--Oseen energy}, and are subject to the vacuum boundary condition $\mathbf{n}(r \rightarrow \infty)=\mathbf{e}_z$.
This boundary condition compactifies the domain $\mathbb{R}^2 \cup\{\infty\} \cong S^2$, such that skyrmions belong to the homotopy group $\pi_2(\mathbb{R}P^2)=\pi_2(S^2)=\mathbb{Z}$.
That is, skyrmions have an associated integer-valued topological degree, given explicitly by
\begin{equation}
    Q_{\textup{Sk}} = \frac{1}{4\pi} \epsilon^{ijk} \int_{\mathbb{R}^2}\textup{d}^2x\, \left(n_i\frac{\partial n_j}{\partial x} \frac{\partial n_k}{\partial y}\right).
\end{equation}

Numerical relaxation of the flexoelectric Frank--Oseen energy \eqref{eq: Flexoelectric Frank--Oseen energy} is carried out using the arrested Newton flow and non-linear conjugate gradient descent methods for the director field $\mathbf{n}:\mathbb{R}^2\rightarrow S^2$ and electric potential $\varphi:\mathbb{R}^2\rightarrow\mathbb{R}$, respectively.
We impose vacuum boundary conditions, $\mathbf{n} \rightarrow\mathbf{e}_z$ and $\varphi\rightarrow0$, as $r\rightarrow\infty$. 


\subsection{Twist favored Bloch skyrmions}
\label{subsec: Twist favored Bloch skyrmions}

In the one constant approximation, the chiral liquid crystal model becomes equivalent to that of a chiral ferromagnetic model with the Dresselhaus DMI term, favoring Bloch modulations.
The Bloch skyrmion ansatz is given by
\begin{equation}
\label{eq: Bloch ansatz}
    \mathbf{n}_{\textup{Bloch}}(r,\theta) = \sin f(r) \mathbf{e}_\theta + \cos f(r) \mathbf{e}_z,
\end{equation}
where $f:\mathbb{R} \rightarrow \mathbb{R}$ is a monotonically decreasing radial profile function with boundary conditions $f(0)=\pi$ and $f(\infty)=0$.
Within this ansatz, the flexoelectric polarization is determined to be be purely radial
\begin{equation}
    \mathbf{P}_{\textup{Bloch}} = \frac{e_3}{e_1}\frac{1}{r}\sin^2 f(r) \mathbf{e}_r.
\end{equation}
While the Bloch ansatz \eqref{eq: Bloch ansatz} is solenoidal, its associated polarization is not,
\begin{equation}
    \bm{\nabla}\cdot\mathbf{P}_{\textup{Bloch}} = \frac{e_3}{e_1} \frac{1}{r}\frac{\textup{d}f}{\textup{d}r} \sin2f(r) \neq 0.
\end{equation}

\begin{figure*}[t]
    \centering
    \begin{subfigure}[b]{\textwidth}
        \includegraphics[width=\textwidth]{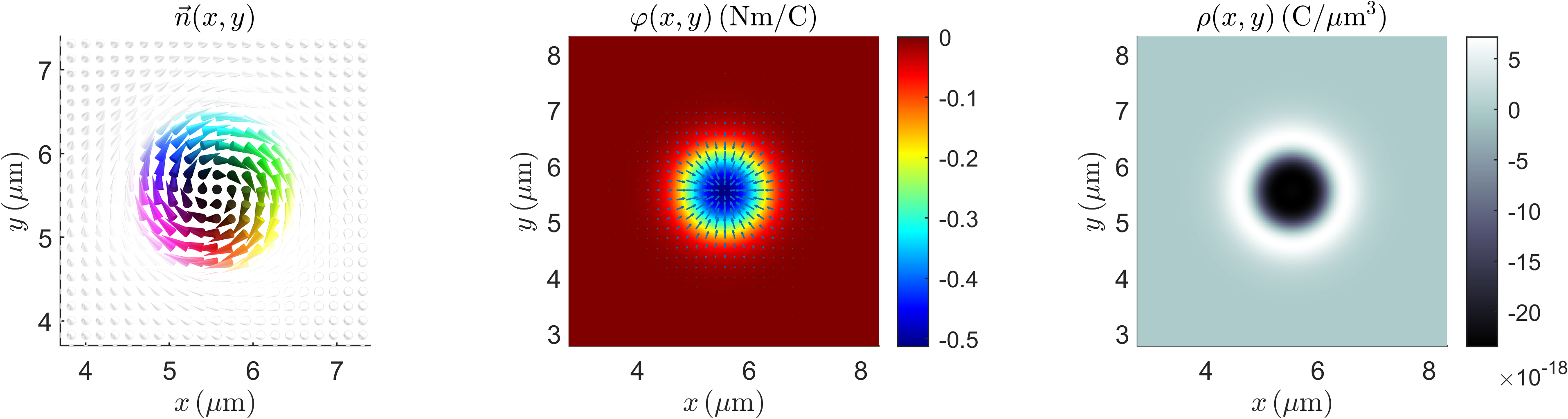}
        \caption{Twist favored Bloch skyrmion}
        \label{fig: DDI results - Bloch skyrmion}
    \end{subfigure}
    \\
    \begin{subfigure}[b]{\textwidth}
        \includegraphics[width=\textwidth]{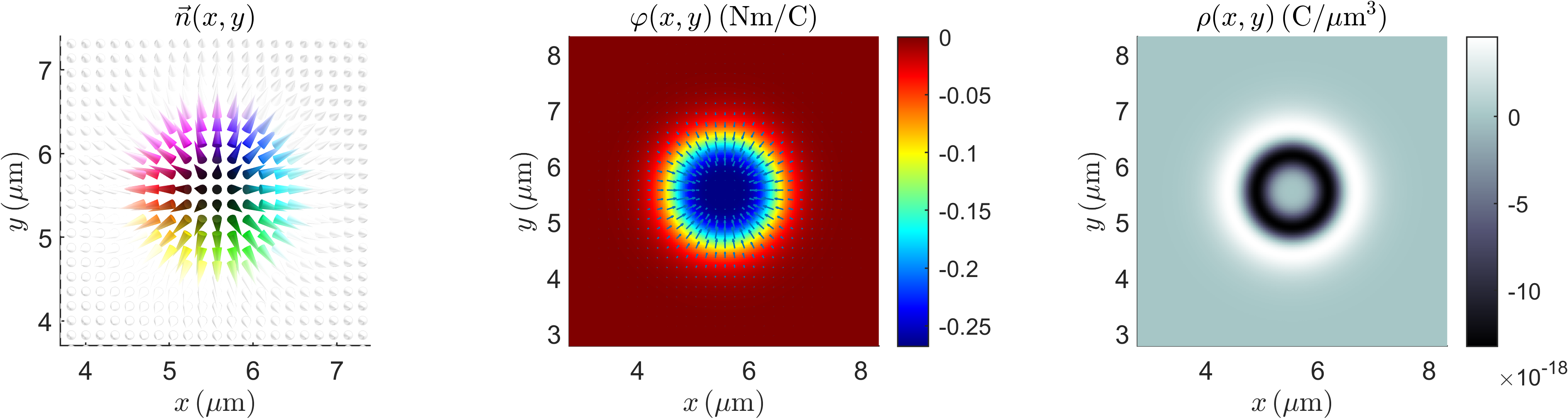}
        \caption{Splay-bend favored N\'eel skyrmion}
        \label{fig: DDI results - Neel skyrmion}
    \end{subfigure}
    \caption{Flexoelectric polarization effect on a skyrmion in a liquid crystal in the one constant approximation. A twist favored Bloch skyrmion is shown in (a), and a splay-bend favored N\'eel skyrmion in (b). In each subfigure is the director $\mathbf{n}(x,y)$ and the flexoelectric potential $\varphi(x,y)$ is shown alongside the electric field vector $\mathbf{E}=-\bm{\nabla}\varphi$. Also shown in is the associated self-induced electric charge density $\rho=-\bm{\nabla}\cdot\mathbf{P}_f$. It can be seen that the Bloch skyrmion has a core of negative charge, surrounded by an outer ring of positive charge. Whereas, the N\'eel skyrmions has a neutral core surrounded by an internal negatively charged ring and a positively charged outer ring. For material properties, we use the elastic coefficient $K=10\,\textup{pN}$ and the cholesteric pitch is chosen to be $p=7\,\mu\textup{m}$. The flexoelectric coefficients are chosen to be $e_1=2\,p\textup{Cm}^{-1}$ and $e_3=4\,p\textup{Cm}^{-1}$, with an applied electric field of strength $E_z=1\,\textup{V}/\mu\textup{m}$ and dielectric anisotropy $\Delta\epsilon=3.7$.}
    \label{fig: DDI results - Skyrmion}
\end{figure*}

Bloch skyrmions in chiral magnets are solenoidal and are unaffected by the magnetostatic self-interaction.
However, Bloch skyrmions in liquid crystals yield non-solenoidal flexoelectric polarization and thus are affected by the electrostatic self-interaction.
This can be seen in Fig.~\ref{fig: DDI results - Bloch skyrmion}.

\subsection{Splay and bend favored N\'eel skyrmions}
\label{subsec: Splay and bend favored Neel skyrmions}

What happens if we now consider liquid crystals which prefer splay and bend, opposed to twist.
Let us remain in the one constant approximation.
Then the Frank--Oseen free energy takes the form
\begin{align}
    F = \, & \frac{K}{2} \int_{\Omega} \textup{d}^3x \left\{ (\mathbf{S}+\mathbf{S}_0)^2 + T^2 + (\mathbf{B}+\mathbf{B}_0)^2 \right\} \nonumber \\
    = \, & \frac{K}{2} \int_{\Omega} \textup{d}^3x  \left\{ (\bm{\nabla}\cdot\mathbf{n})^2 + 2 \mathbf{S}_0\cdot\mathbf{n}(\bm{\nabla}\cdot\mathbf{n}) + [\mathbf{n} \cdot (\bm{\nabla} \times \mathbf{n})]^2 \right. \nonumber \\
    \, &  \left.  + [\mathbf{n} \times (\bm{\nabla} \times \mathbf{n})]^2 - 2 \mathbf{B}_0\cdot[(\mathbf{n}\cdot\bm{\nabla})\mathbf{n}] +\textup{const.}\right\}.
\end{align}
If we choose $\mathbf{S}_0=\mathbf{B}_0=q_0\mathbf{e}_z$, then the model reduces to that of the chiral magnet with the DMI term arising from the Rashba SOC.
That is, the free energy becomes
\begin{align}
    F = \, & \int_{\Omega} \textup{d}^3x  \left\{ \frac{K}{2}(\nabla\mathbf{n})^2 + Kq_0 [n_z(\bm{\nabla}\cdot \mathbf{n}) - \mathbf{n}\cdot \bm{\nabla}n_z] \right. \nonumber \\
    \, &  \left. + V(\mathbf{n}) \right\},
\end{align}
where we have included the potential term and used the identity \eqref{eq: Gradient identity} again.
The factor of $q_0$ was chosen for convenience as we can pick the same energy and length scales as the twist favored model.
It also allows us to compare splay-bend favored N\'eel skyrmions with the twist favored Bloch skyrmions.
For more insight into the various DMI types and their associated skyrmions, see e.g. \cite{Gobel_2021}.

Let us now include the electrostatic self-energy, and employ the same length $L_0=1/q_0 $ and energy $E_0=K/q_0$ scales as before.
Then, in the translation invariant case, the normalized free energy of this splay-bend favored liquid crystal model becomes
\begin{align}
    F = \, & \int_{\mathbb{R}^2} \textup{d}^2x \left\{ \frac{1}{2} (\nabla \mathbf{n})^2 + \left[ n_z(\bm{\nabla}\cdot \mathbf{n}) - \mathbf{n}\cdot \bm{\nabla}n_z \right]  \right.\nonumber \\
    \, & \left. + \frac{1}{q_0^2} \frac{1}{K} V(\mathbf{n}) + \frac{\epsilon}{2}\varphi\Delta\varphi \right\}.
\end{align}
It is well-known that the Rashba DMI term prefers N\'eel hedgehog skyrmions, given by the ansatz
\begin{equation}
\label{eq: Neel ansatz}
    \mathbf{n}_{\textup{N\'eel}}(r,\theta) = \sin f(r) \mathbf{e}_r + \cos f(r) \mathbf{e}_z.
\end{equation}
The self-induced polarization, coming from the N\'eel ansatz \eqref{eq: Neel ansatz}, is
\begin{align}
    &\mathbf{P}_{\textup{N\'eel}} =  \left[\frac{1}{r}\sin^2f(r) + \left(1-\frac{e_3}{e_1}\right)\frac{1}{2}\sin2f(r) \frac{\textup{d}f}{\textup{d}r}\right] \mathbf{e}_r \nonumber \\
    \, & + \left[ \frac{1}{2r}\sin2f(r) + \left( \cos^2f(r) + \frac{e_3}{e_1}\sin^2f(r) \right) \frac{\textup{d}f}{\textup{d}r} \right] \mathbf{e}_z.
\end{align}
Unlike the Bloch polarization, the N\'eel polarization picks up an out-of-plane component.
The divergence of this polarization is also non-zero.
We note that if the flexoelectric coefficients are equal, $e_1=e_3$, then the divergence of the polarization for both Bloch and N\'eel ans\"atze are the same,
\begin{equation}
    \bm{\nabla}\cdot\mathbf{P}_{\textup{Bloch}} = \bm{\nabla}\cdot\mathbf{P}_{\textup{N\'eel}} = \frac{1}{r}\frac{\textup{d}f}{\textup{d}r} \sin2f(r).
\end{equation}
So, they will generate the same electrostatic potential and, thus, they will be energy degenerate for equal flexoelectric coefficients.
For unequal flexoelectric coefficients, $e_1 \neq e_3$, the associated self-induced polarizations yield different electric scalar potentials and, hence, distinct skyrmions. 
This can be seen in Fig.~\ref{fig: DDI results - Neel skyrmion}.


\section{Liquid crystal hopfions}
\label{sec: Liquid crystal hopfions}

\begin{figure*}[t]
    \centering
    \begin{subfigure}[b]{0.4\textwidth}
        \includegraphics[width=\textwidth]{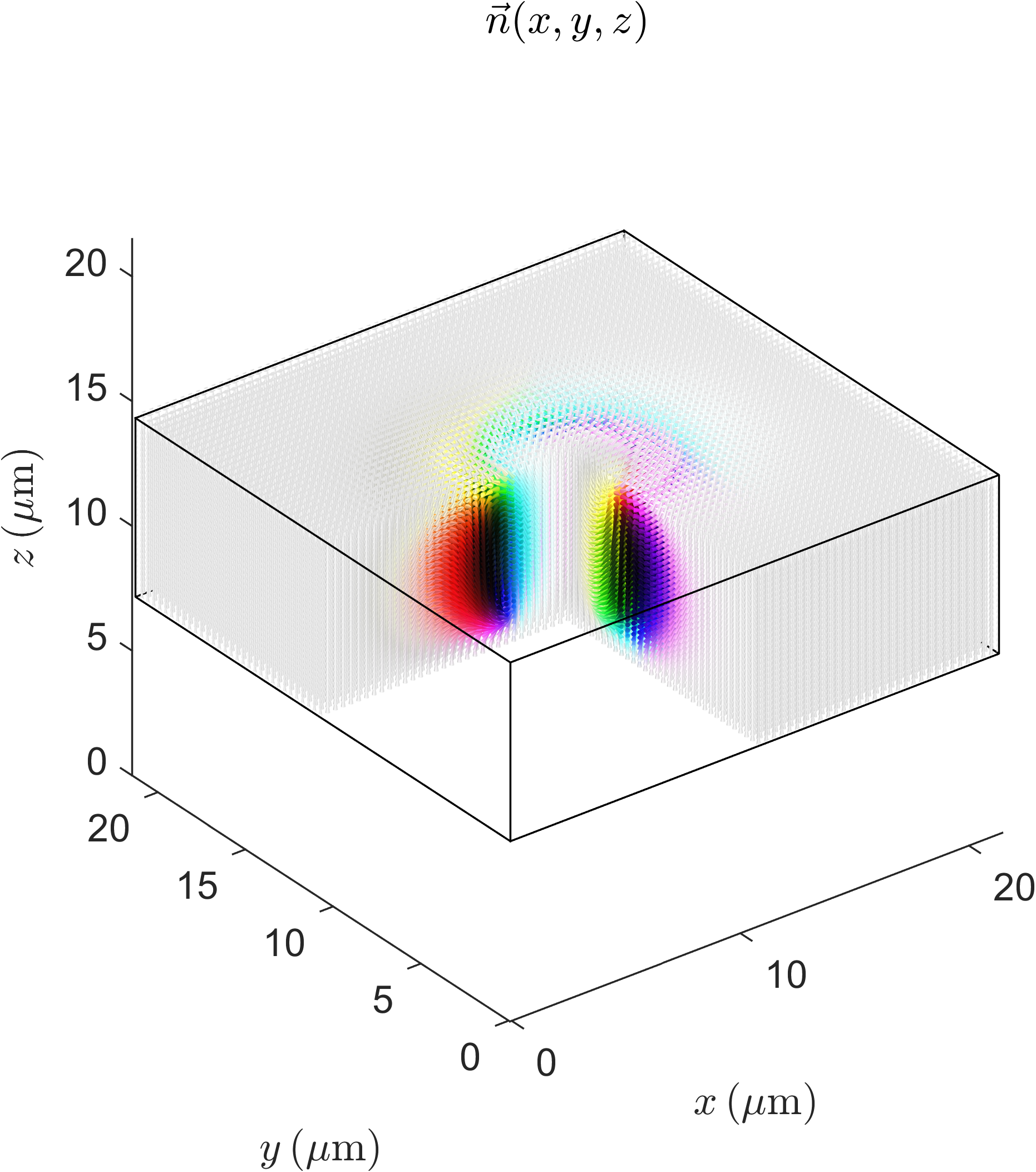}
    \end{subfigure}
    ~
    \begin{subfigure}[b]{0.4\textwidth}
        \includegraphics[width=\textwidth]{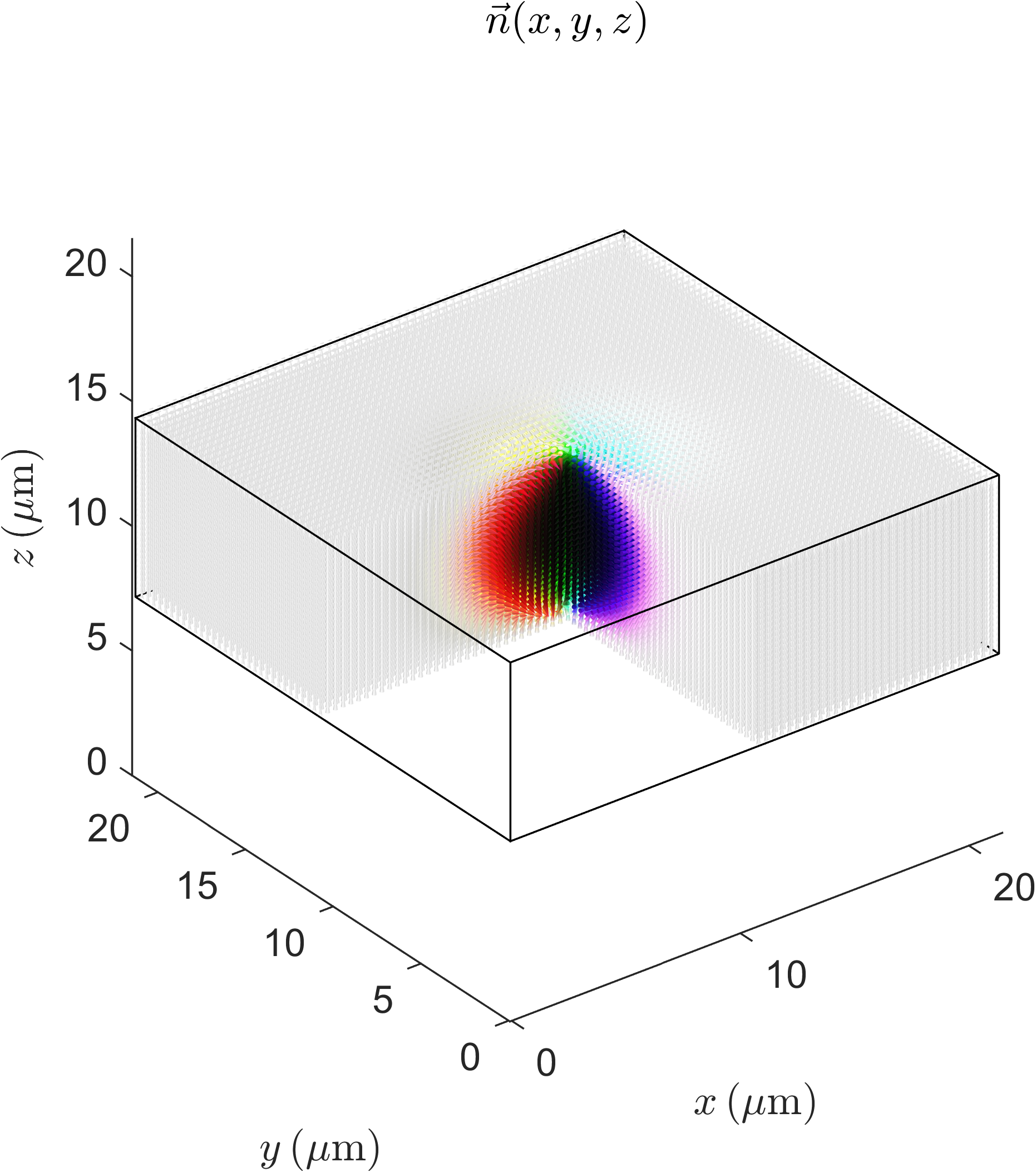}
    \end{subfigure}
    \\
    \begin{subfigure}[b]{0.4\textwidth}
        \includegraphics[width=\textwidth]{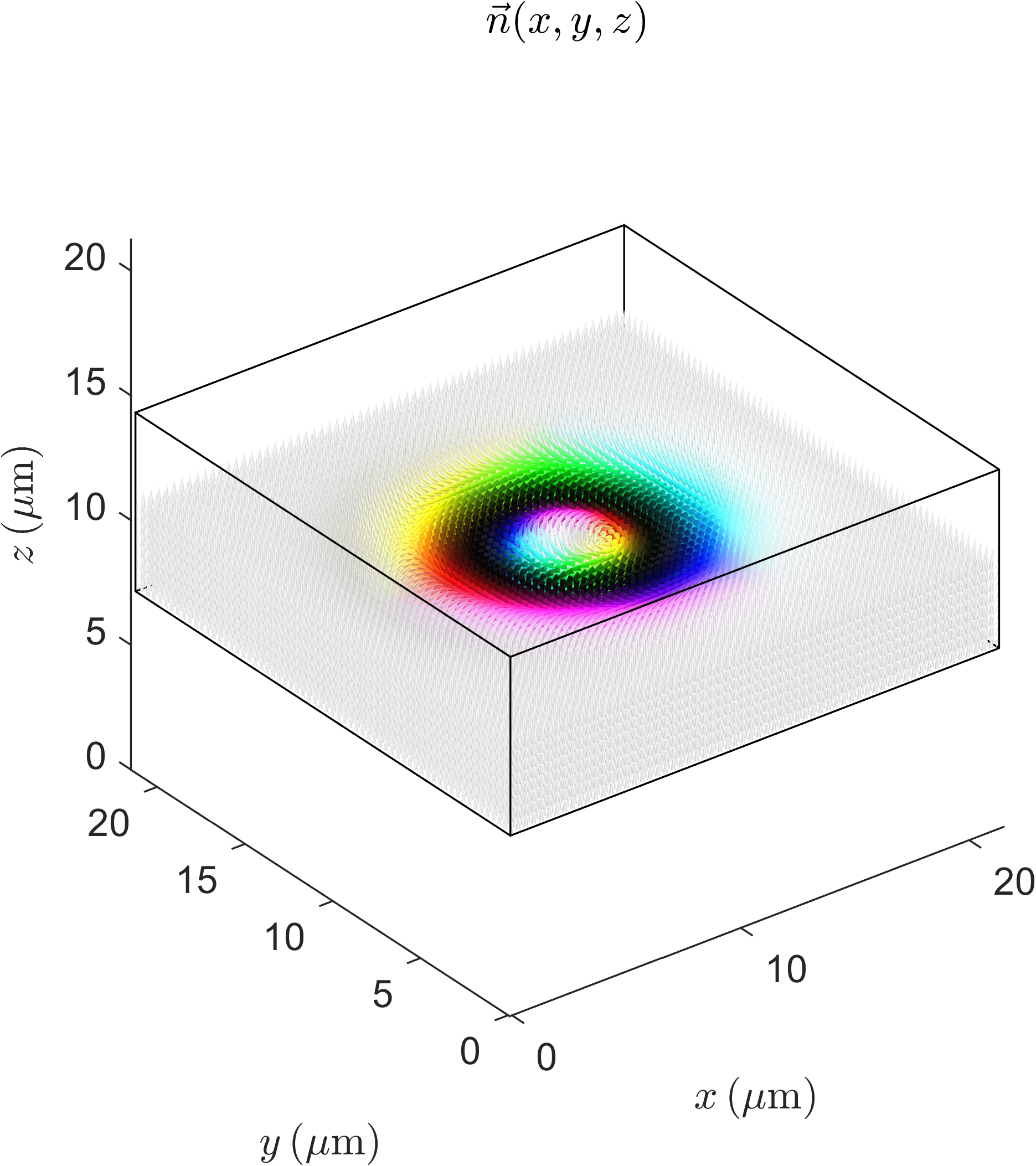}
        \caption{Hopfion ($e_1=e_3=4\,\textup{pCm}^{-1}$)}
        \label{fig: Director field - Hopfion}
    \end{subfigure}
    ~
    \begin{subfigure}[b]{0.4\textwidth}
        \includegraphics[width=\textwidth]{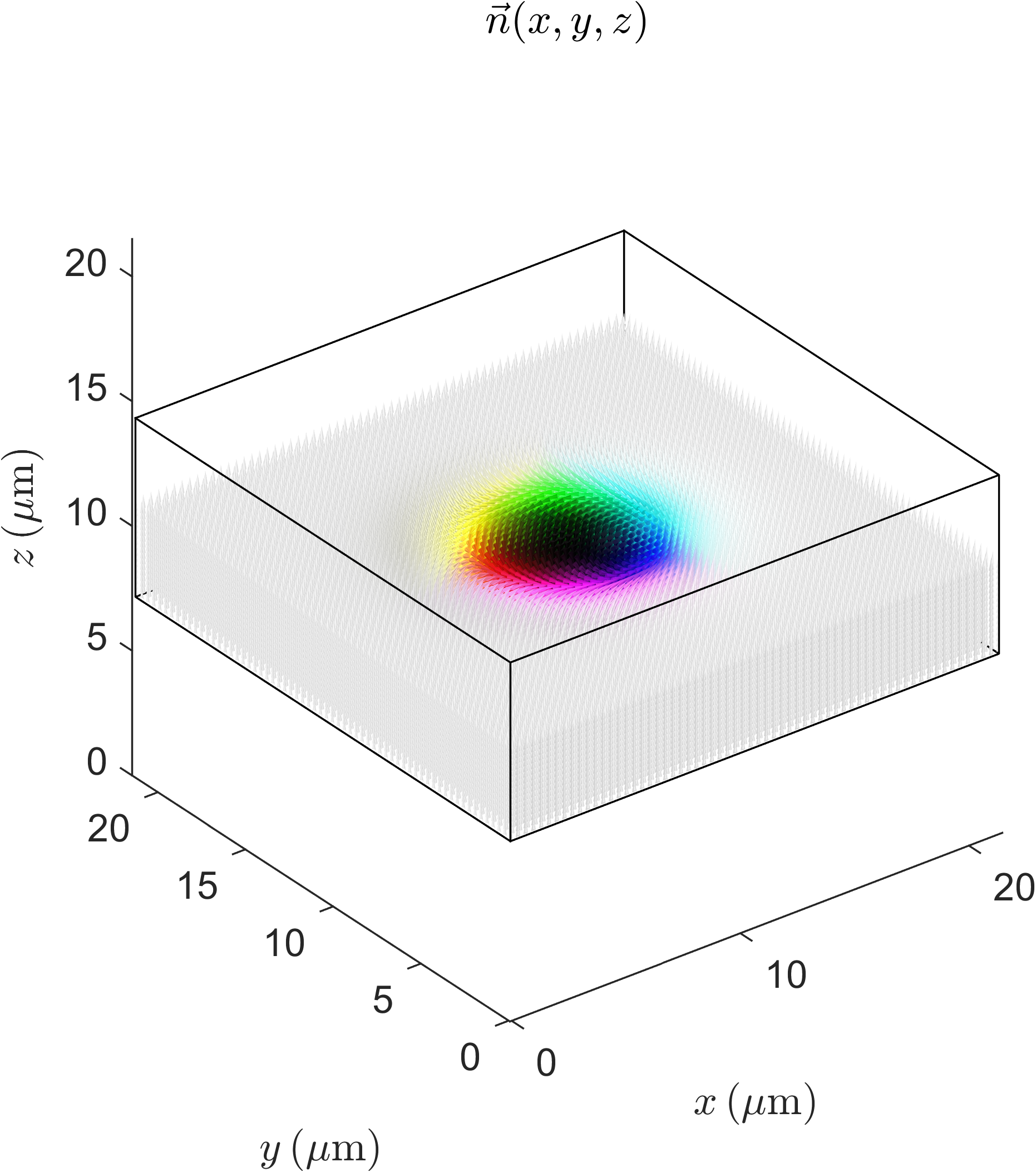}
        \caption{toron ($e_1=e_3=8\,\textup{pCm}^{-1}$)}
        \label{fig: Director field - Skyrmion}
    \end{subfigure}
    \caption{Flexoelectric polarization effect on a hopfion in a liquid crystal in the one constant approximation $K_i=K=10\,\textup{pN}$. Shown is the relaxed director field $\mathbf{n}(x,y,z)$ solution of the flexoelectric Frank--Oseen energy \eqref{eq: Flexoelectric Frank--Oseen energy}, starting from the hopfion ansatz \eqref{eq: Hopfion ansatz} as an initial configuration. Plots of the director field $\mathbf{n}$ are shown for (a) a hopfion and (b) a toron. The distance between the cell plates is taken to be $d=7\,\mu\textup{m}$, with an applied voltage of $U=2\,V$. This gives an applied electric field strength of $E_z=U/d=0.29\,\textup{V}/\mu\textup{m}$ and we set the dielectric anisotropy to be $\Delta\epsilon=4.8$. The cholesteric pitch $p$ is taken to be the same length as the distance between the cell plates, $p=d$. We set the flexoelectric coefficients to be equal $e_1=e_3$. There is a phase transition from a hopfion to a toron as the flexoelectric coefficients are increased from $e_i=4\,\textup{pCm}^{-1}$ to $e_i=8\,\textup{pCm}^{-1}$.}
    \label{fig: Hopfion results - Director field}
\end{figure*}

A hopfion can be interpreted as a twisted skyrmion string, forming a closed loop in real space.
Cross-sectional views (top panel of Fig.~\ref{fig: Director field - Hopfion}) show the skyrmion twisting as it winds around the hopfion core, changing from an in-plane skyrmion ($Q_{\textup{Sk}} =-1$) to an out-of-plane antiskyrmion ($Q_{\textup{Sk}} =+1$).
A cross-section view through the center of the disk ($z=0$) gives a different perspective of the hopfion, showing the structure of Bloch skyrmionium ($Q_{\textup{Sk}} =0$), or a $2\pi$-vortex \cite{Bogdanov_1999} (bottom panel of Fig.~\ref{fig: Director field - Hopfion}).

Hopfions comprise inter-linked closed-loop preimages of constant $\mathbf{n}(x,y,z)$ \cite{Tai_2018}.
The linking of closed-loop preimages of anti-podal points in $S^2/\mathbb{Z}_2 \cong \mathbb{R}P^2$ defines a homotopy invariant.
This homotopy invariant is the Hopf index $Q_{\textup{Hopf}}\in\pi_3(\mathbb{R}P^2)=\pi_3(S^2)=\mathbb{Z}$, associated to the fibration $\mathbb{Z}_2\rightarrow S^2 \rightarrow\mathbb{R}P^2$.
The Hopf index of a given vector field $\mathbf{F}$ can be computed by constructing a gauge potential $\mathbf{A}$, such that $\mathbf{F}=\bm{\nabla}\times\mathbf{A}$ and \cite{Knapman_2025,Souza_2025}
\begin{equation}
    F_i = \frac{1}{8\pi^2}\epsilon_{ijk}\mathbf{n} \cdot (\partial_j\mathbf{n} \times \partial_k \mathbf{n}).
\end{equation}
Then the Hopf index is given by \cite{Hietarinta_2012}
\begin{equation}
    Q_{\textup{Hopf}} = -\int_{\Omega} \textup{d}^3x \, \mathbf{F}\cdot\mathbf{A}.
\end{equation}
We will consider a hopfion with Hopf index $Q_{\textup{Hopf}}=1$, defined by the ansatz \cite{Hietarinta_1999}
\begin{equation}
\label{eq: Hopfion ansatz}
    \mathbf{n}_{\textup{Hopf}} =
    \begin{pmatrix}
        \frac{4\Sigma r\left(\Theta \cos\theta - (\Lambda-1)\sin\theta \right)}{(1+\Lambda)^2} \\
        \frac{4\Sigma r\left(\Theta \sin\theta + (\Lambda-1)\cos\theta \right)}{(1+\Lambda)^2} \\
        1 - \frac{8\Sigma^2r^2}{(1+\Lambda)^2}
    \end{pmatrix},
\end{equation}
where we have introduced the three functions \cite{Sutcliffe_2018}
\begin{subequations}
  \begin{align}
    \Theta(z) = \, & \tan\left(\frac{\pi z}{d}\right), \\
    \Sigma(r,z) = \, & \frac{1}{d}\left[ 1 + \left(\frac{2z}{d}\right)^2 \right] \sec\left( \frac{\pi r}{2d} \right), \\
    \Lambda(r,z) = \, & \Sigma^2r^2+\frac{\Theta^2}{4}.
\end{align}  
\end{subequations}

\begin{figure*}[t]
    \centering
    \begin{subfigure}[b]{0.4\textwidth}
        \includegraphics[width=\textwidth]{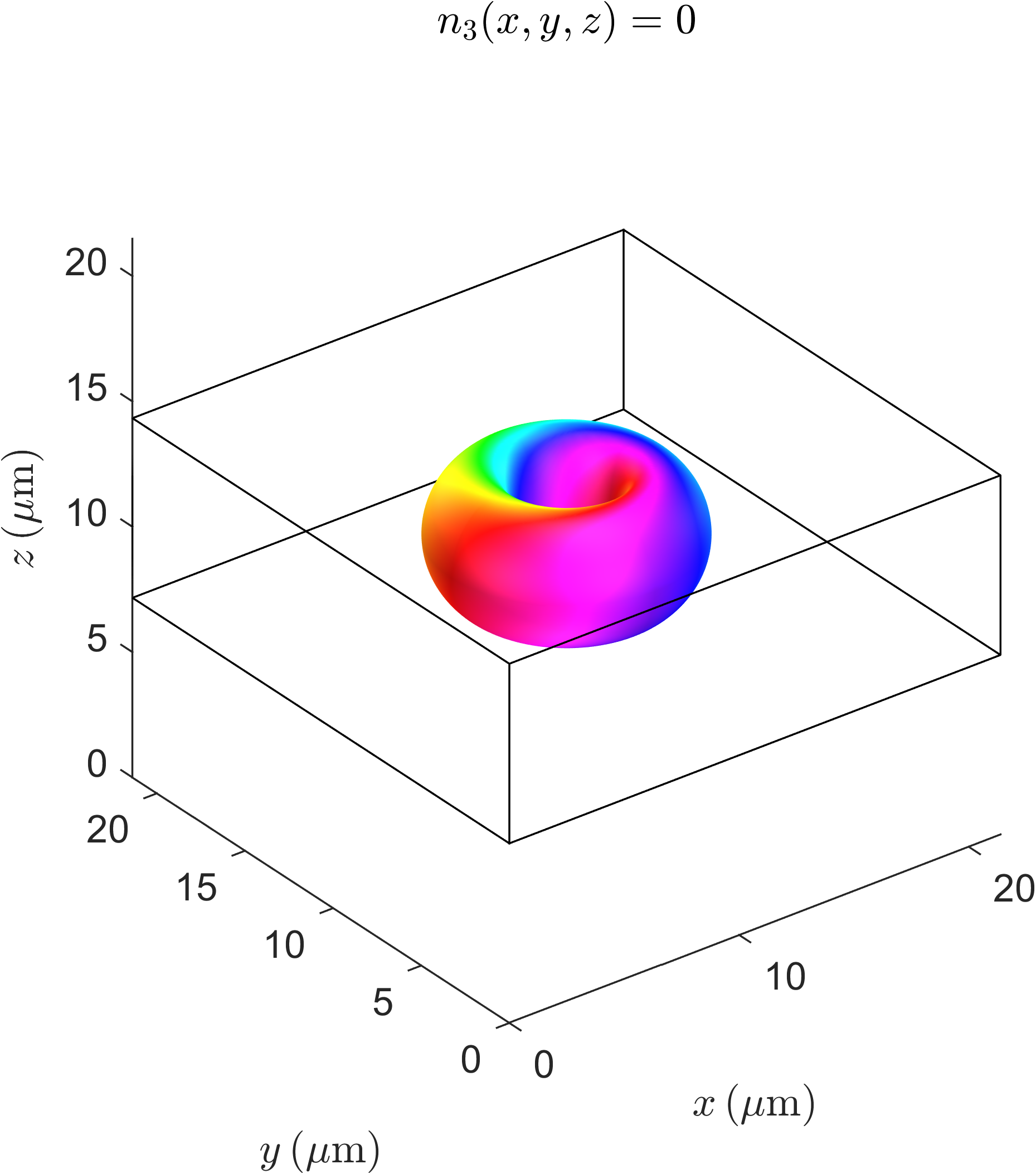}
    \end{subfigure}
    ~
    \begin{subfigure}[b]{0.4\textwidth}
        \includegraphics[width=\textwidth]{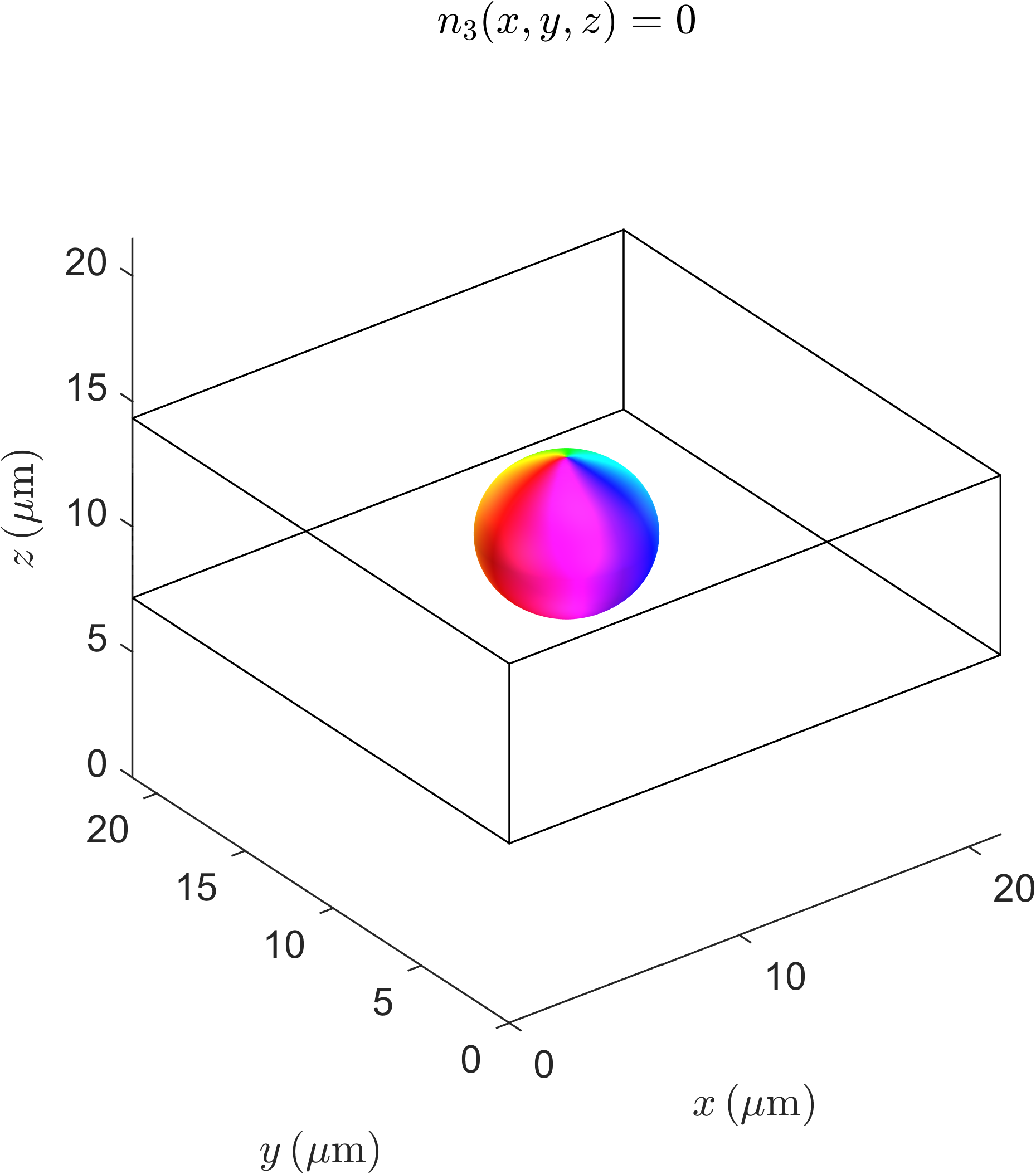}
    \end{subfigure}
    \\
    \begin{subfigure}[b]{0.4\textwidth}
        \includegraphics[width=\textwidth]{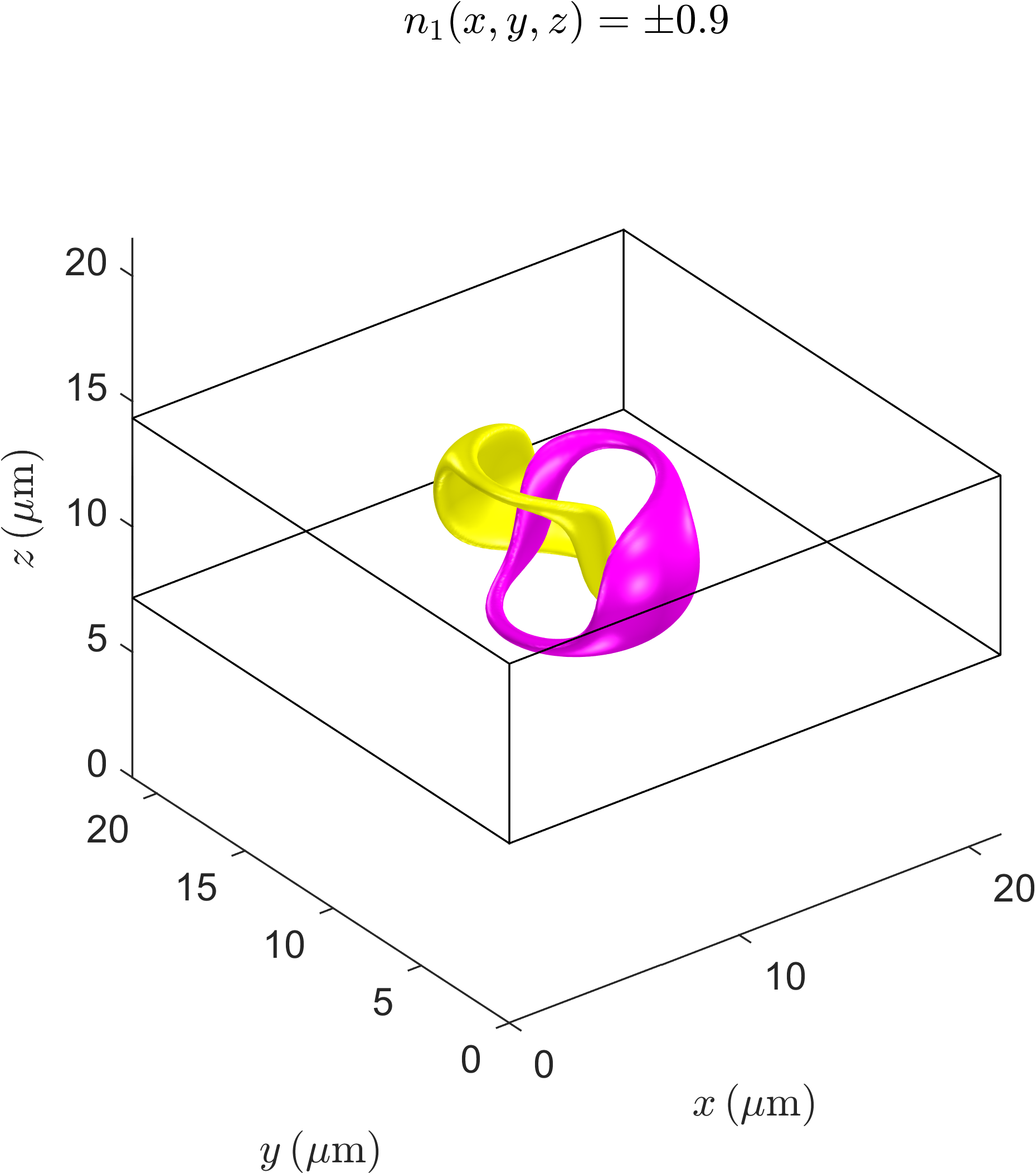}
        \caption{Hopfion ($e_1=e_3=4\,\textup{pCm}^{-1}$)}
        \label{fig: Linking hopfion}
    \end{subfigure}
    ~
    \begin{subfigure}[b]{0.4\textwidth}
        \includegraphics[width=\textwidth]{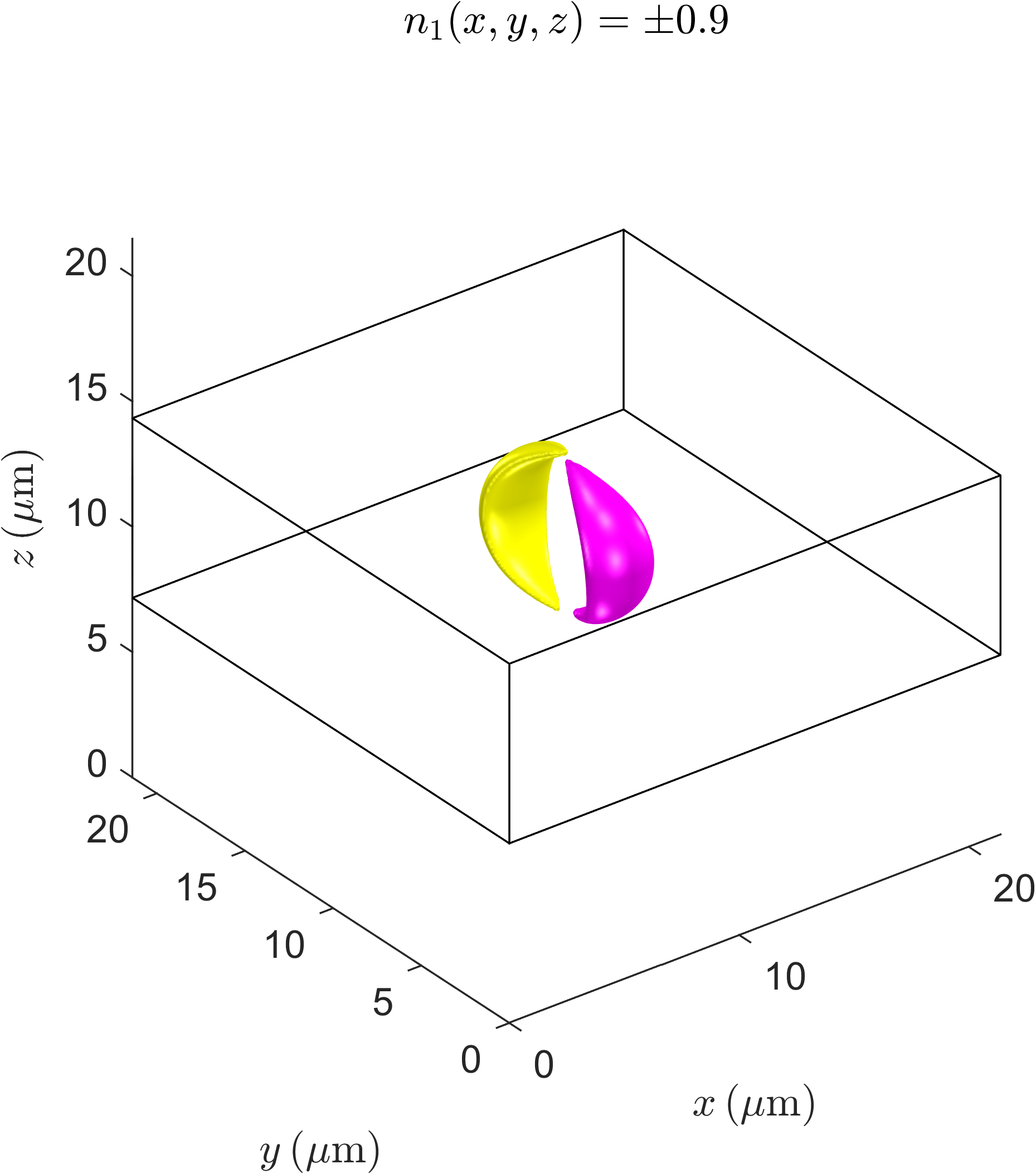}
        \caption{toron ($e_1=e_3=8\,\textup{pCm}^{-1}$)}
        \label{fig: Linking skyrmion}
    \end{subfigure}
    \caption{Isosurface plots of the director field: the top row is the field $n_3=0$ and the bottom row is the linking of the field $n_1=\pm0.9$. The hopfion is shown in (a) and the toron in (b). The parameter set is detailed in Figure \ref{fig: Hopfion results - Director field}. As the flexoelectric coefficients are increased from $e_i=4\,\textup{pCm}^{-1}$ to $e_i=8\,\textup{pCm}^{-1}$, the hopfion transitions to a toron. It is clear to see from the bottom row that the knots become unlinked.}
    \label{fig: Hopfion results - Fields and linking}
\end{figure*}

Using the hopfion ansatz \eqref{eq: Hopfion ansatz} as an initial configuration for our numerical minimization algorithm, the relaxed director solution is plotted in Fig.~\ref{fig: Director field - Hopfion}.
Consider the closed-loop preimages of $\mathbf{n}=(1,0,0)$ and $\mathbf{n}=(-1,0,0)$ by letting us construct two isosurface tubes, each corresponding to closed-loop preimages about $\mathbf{n}=(\pm1,0,0)$.
Such a construction is shown in Fig.~\ref{fig: Linking hopfion}.
By considering the preimages of $\mathbf{n}=(\pm1,0,0)$, it is clear to see they do indeed only link once.

In our simulations we use a grid with lattice points $n_x=n_y=128$ and $n_z=192$ to model $\mathbb{R}^3$.
This corresponds to a discretised model of the disk $\Omega$ with lattice points $128\times128\times64$.
We add a layer of vacuum above and below the disk, with the same thickness as the disk, $d$.
Different vacuum layer thicknesses were tried but a vacuum layer with thickness of $O(d)$ was sufficiently accurate.
For the electric scalar potential $\varphi:\mathbb{R}^3 \rightarrow\mathbb{R}$, we employ a central finite difference method throughout $\mathbb{R}^3$.
However, for the director field $\mathbf{n}:\Omega\rightarrow\mathbb{R}P^2$, we use a central finite difference method in the bulk of the disk $\Omega$ and a backward/forward finite difference method on the disk boundary $\partial\Omega$.

\begin{figure*}[t]
    \centering
    \begin{subfigure}[b]{0.415\textwidth}
        \includegraphics[width=\textwidth]{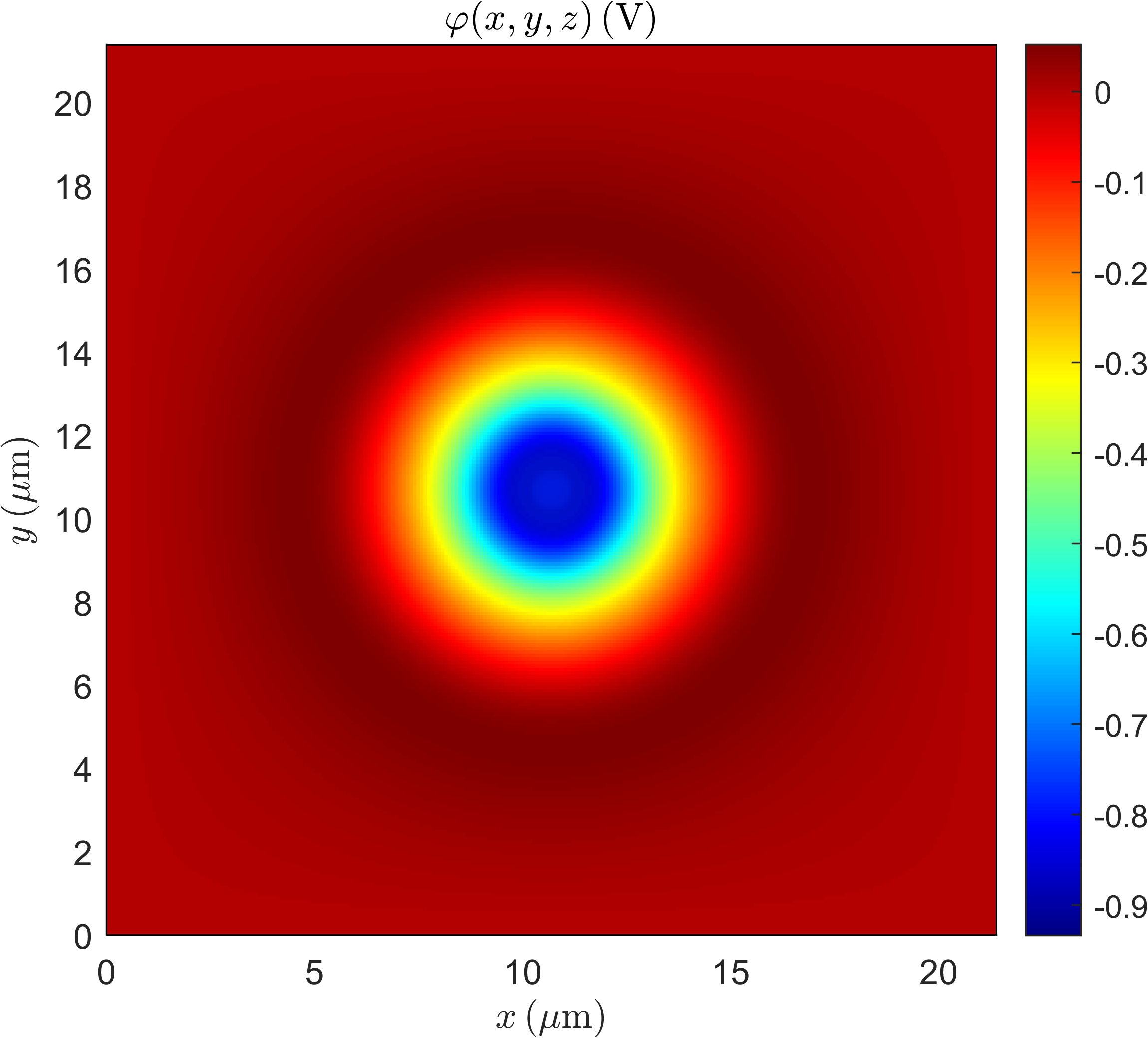}
    \end{subfigure}
    ~
    \begin{subfigure}[b]{0.45\textwidth}
        \includegraphics[width=\textwidth]{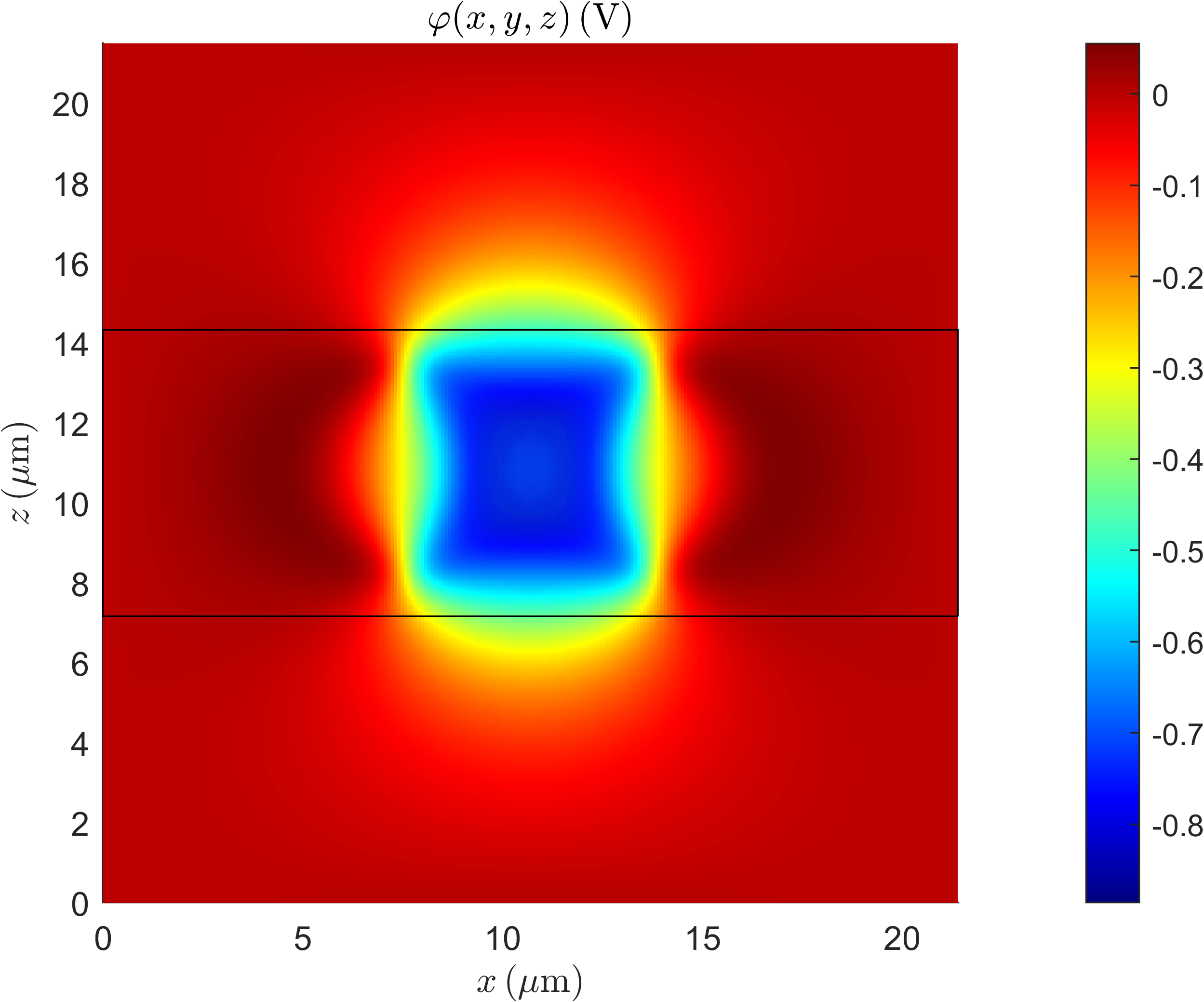}
    \end{subfigure}
    \\
    \begin{subfigure}[b]{0.415\textwidth}
        \includegraphics[width=\textwidth]{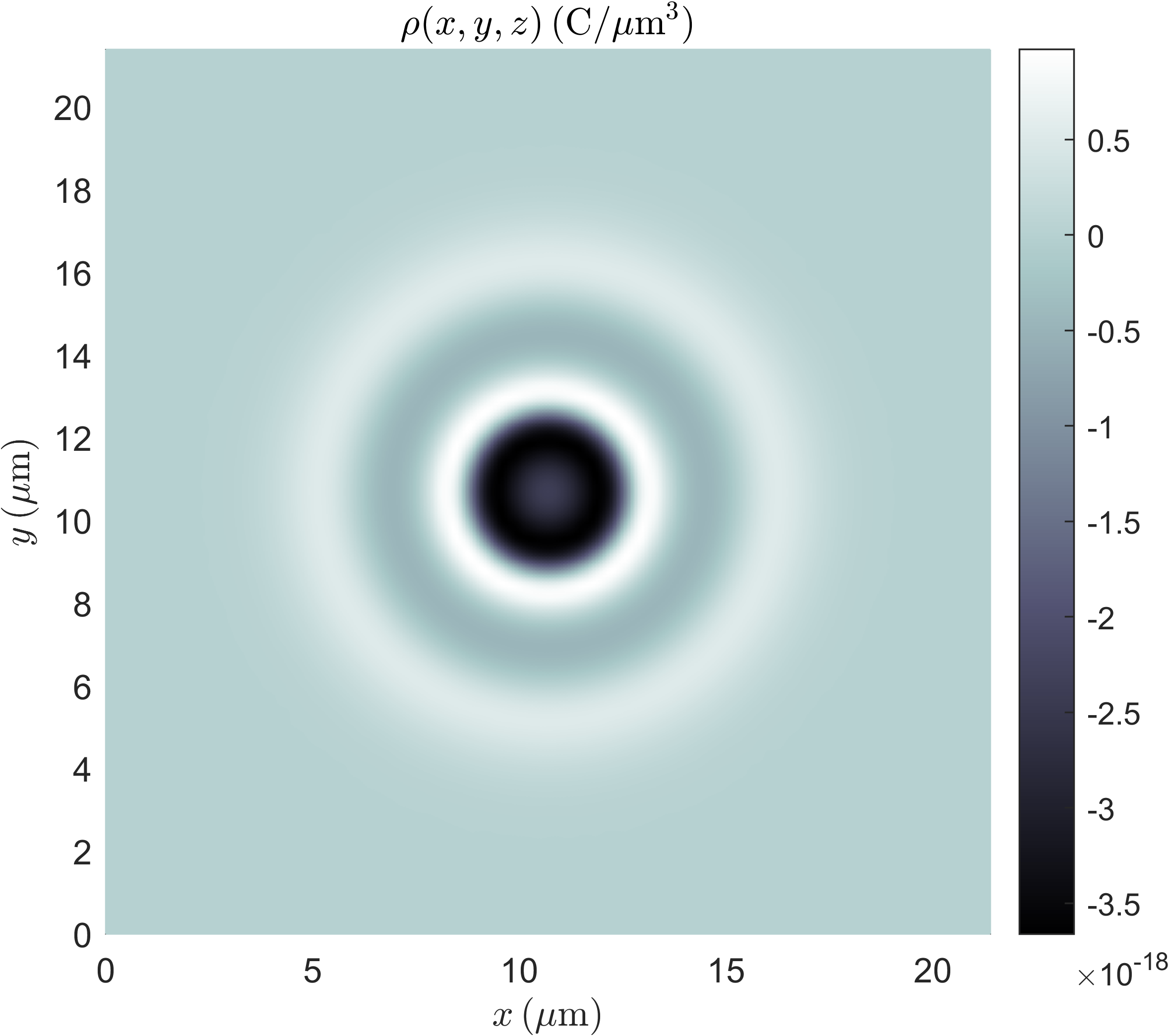}
    \end{subfigure}
    ~
    \begin{subfigure}[b]{0.45\textwidth}
        \includegraphics[width=\textwidth]{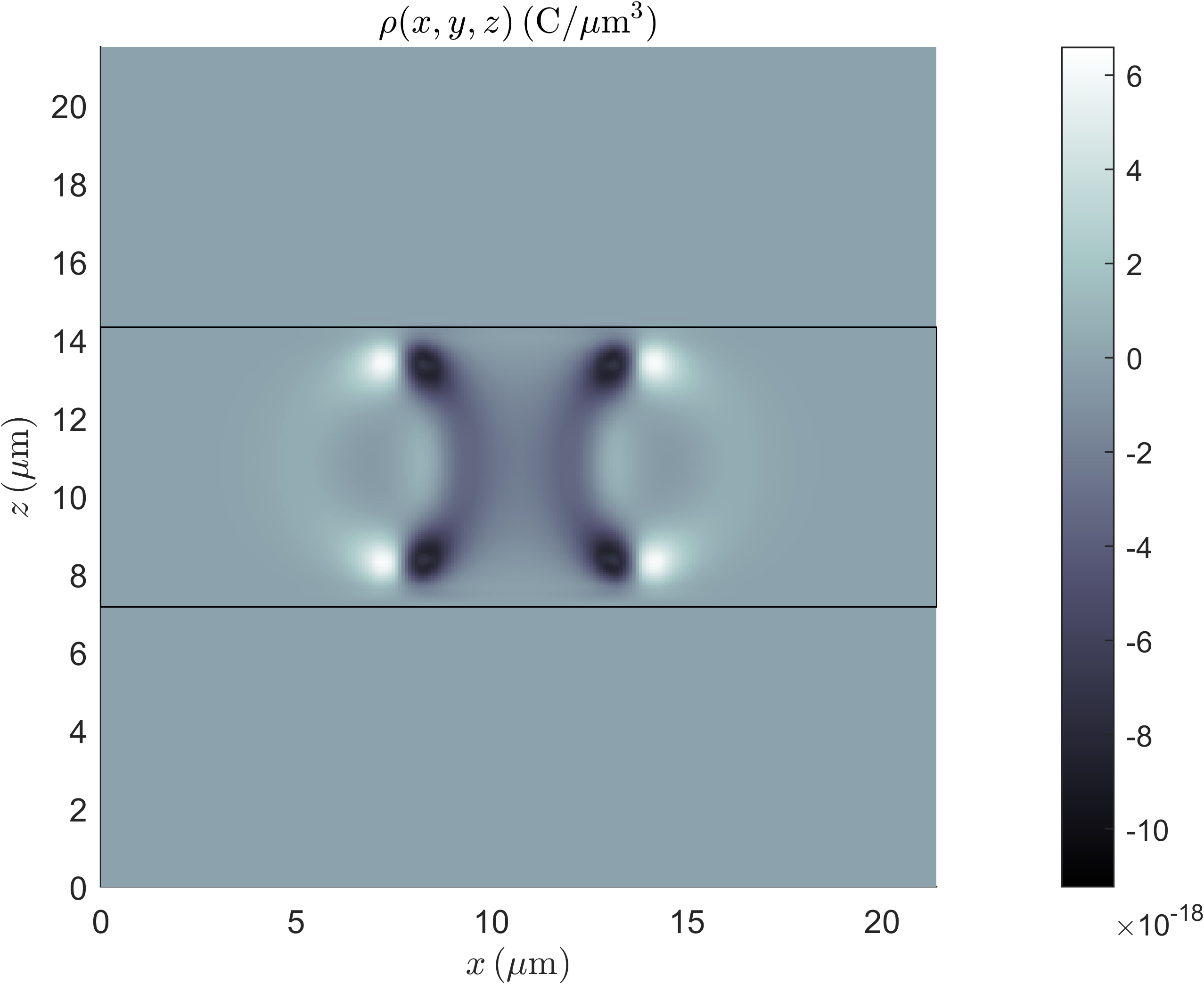}
    \end{subfigure}
    \caption{Flexoelectric polarization effect on a hopfion in a liquid crystal in the one constant approximation. The top row shows the self-induced electrostatic scalar potential $\varphi$, which is a solution of the Poisson equation $\Delta\varphi=\rho/\epsilon_0$, and the bottom row displays the electric charge density $\rho=-\bm{\nabla}\cdot\mathbf{P}_f$. It is clear to see that the electric charge density is confined within the confined geometry $\Omega$, whereas the electric scalar potential extends into $\mathbb{R}^3/\Omega$. The parameter set is detailed in Figure \ref{fig: Hopfion results - Director field} with flexoelectric coefficients $e_1=e_3=4\,\textup{pCm}^{-1}$.}
    \label{fig: Hopfion results - Electric properties}
\end{figure*}

The resulting electrostatic properties of the minimal energy $Q_{\textup{Hopf}}=1$ hopfion, associated to the flexoelectric Frank--Oseen energy \eqref{eq: Flexoelectric Frank--Oseen energy}, are plotted in Fig.~\ref{fig: Hopfion results - Electric properties} for the hopfion.
We use similar material properties as in the two-dimensional skyrmion case.
This is carried out within the one constant approximation with $K=10\,\textup{pN}$.
We set the distance between the cell plates to be $d=7\,\mu\textup{m}$, with an applied potential difference of $U=2\,\textup{V}$ between the plates.
This generates an electric field of strength $E_z=U/d=0.29\,\textup{V}/\mu\textup{m}$ and the dielectric anisotropy constant is chosen to be $\Delta\epsilon=4.8$.
These parameters are similar to those employed in \cite{Ackerman_2017}.
The flexoelectric constants are normally estimated to be of the order of magnitude $\sim3\,\textup{pCm}^{-1}$ \cite{Blinov_2017} and for nematic liquid crystals in the range $1-10\,\textup{pCm}^{-1}$ \cite{Harden_2006}.
So we choose to set them to initially be equal with $e_1=e_3=4\,\textup{pCm}^{-1}$.
We set our box size for $\mathbb{R}^3$ to be $l_x=l_y=l_z=3 d$ and choose the cholesteric pitch to be $p/d=1$.
Hence, the lattice spacing is $\Delta l_x=\Delta l_y=0.165\,\mu\textup{m}$ and $\Delta l_z=0.110\,\mu\textup{m}$.

The electric scalar potential $\varphi:\mathbb{R}^3\rightarrow\mathbb{R}$ and the electric charge density $\rho:\mathbb{R}^3\rightarrow\mathbb{R}$ are shown in Fig.~\ref{fig: Hopfion results - Electric properties}.
The left column shows a cross-section view on the $xy$-plane at the center of the disk $\Omega$.
(In our numerical simulations this corresponds to $z=3d/2$.)
This view shows the skyrmionium structure and it is clear to see that both the electric potential and charge density (in this cross-section of the hopfion) behave similar to the two-dimensional skyrmion.
A cross-section view on the $xz$-plane at $y=3d/2$ is displayed in the right column of Fig.~\ref{fig: Hopfion results - Electric properties}, showing the axial symmetry of the electric potential and charge density.

\begin{figure*}[t]
    \centering
    \begin{subfigure}[b]{0.415\textwidth}
        \includegraphics[width=\textwidth]{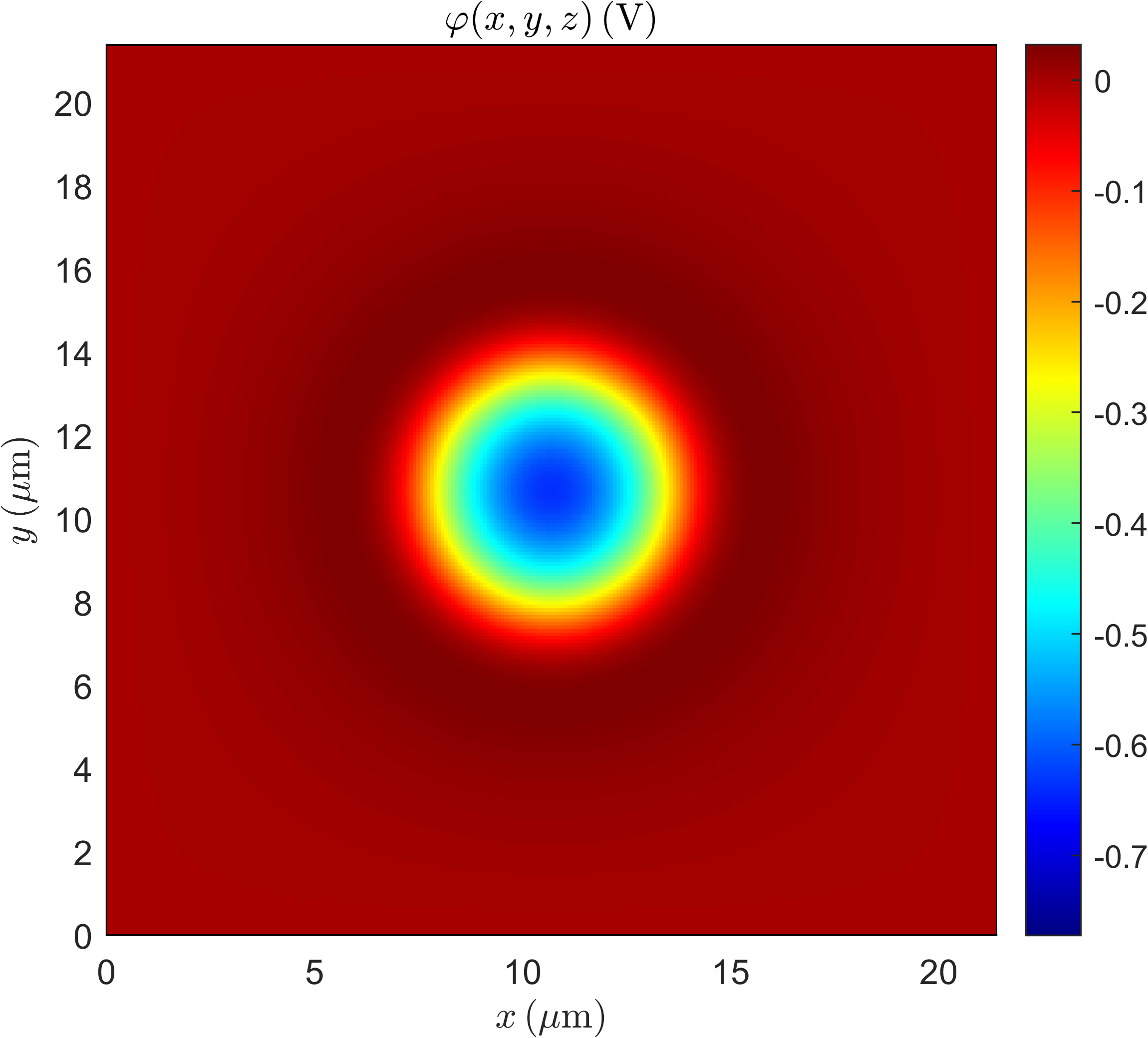}
    \end{subfigure}
    ~
    \begin{subfigure}[b]{0.45\textwidth}
        \includegraphics[width=\textwidth]{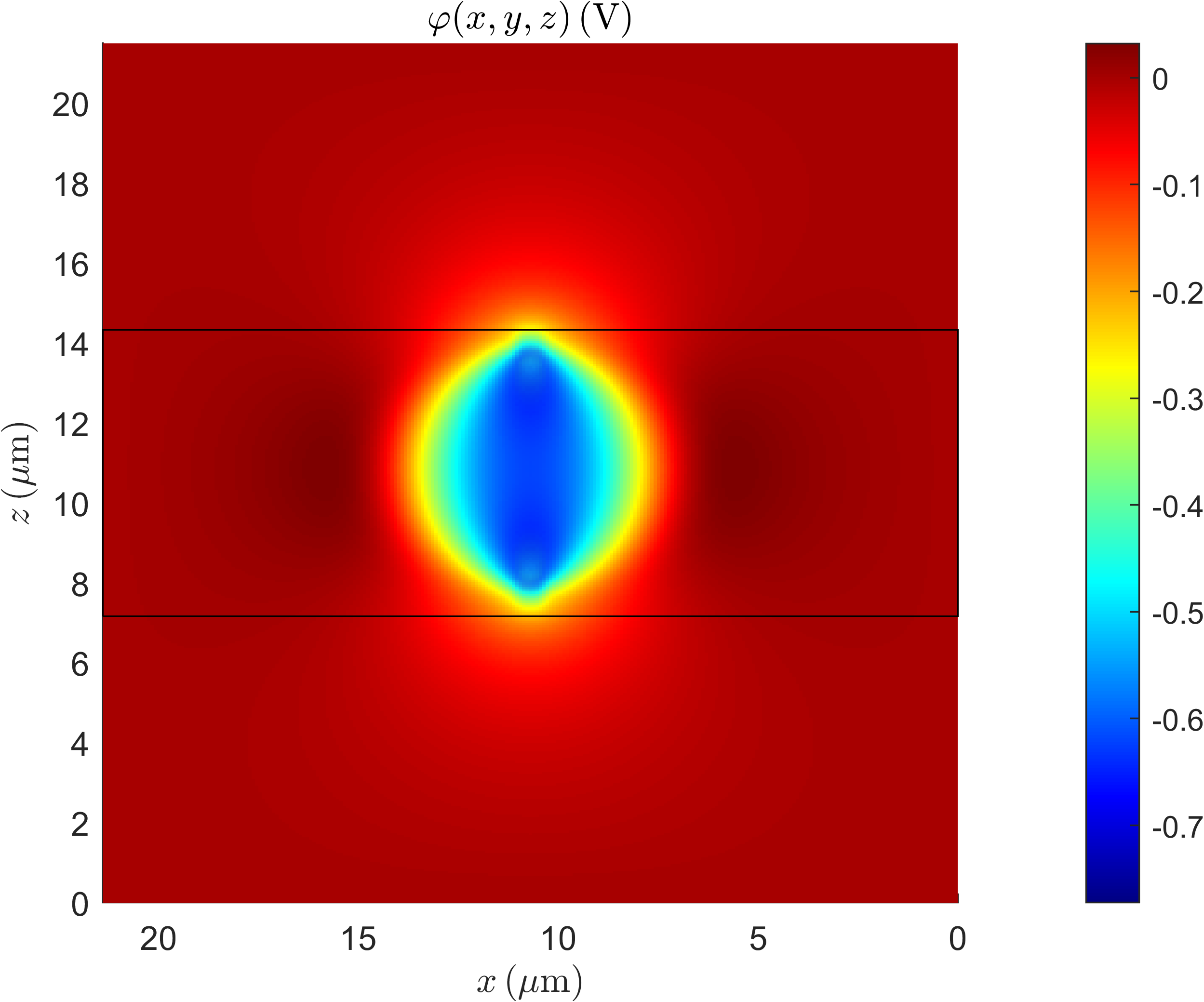}
    \end{subfigure}
    \\
    \begin{subfigure}[b]{0.415\textwidth}
        \includegraphics[width=\textwidth]{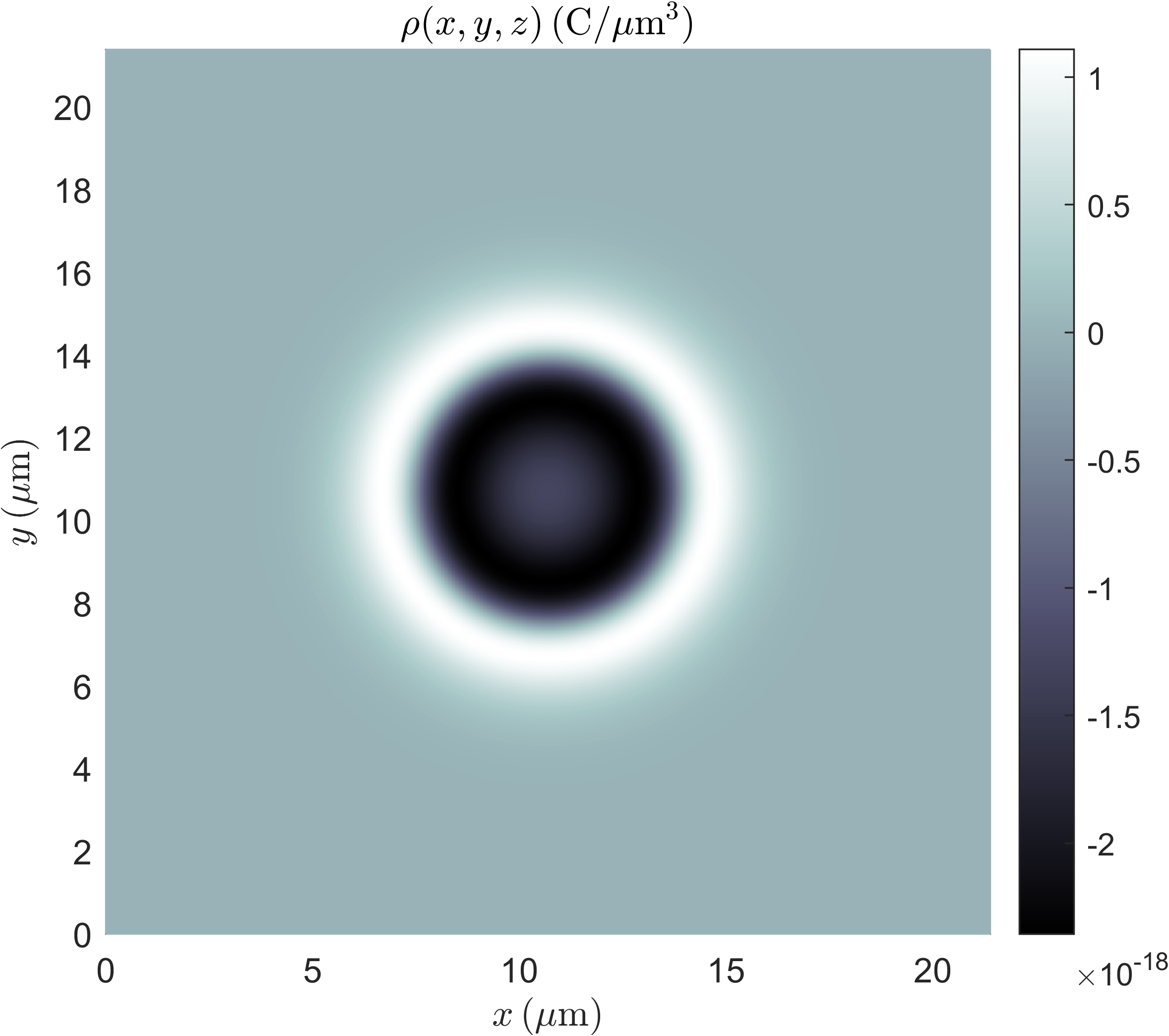}
    \end{subfigure}
    ~
    \begin{subfigure}[b]{0.45\textwidth}
        \includegraphics[width=\textwidth]{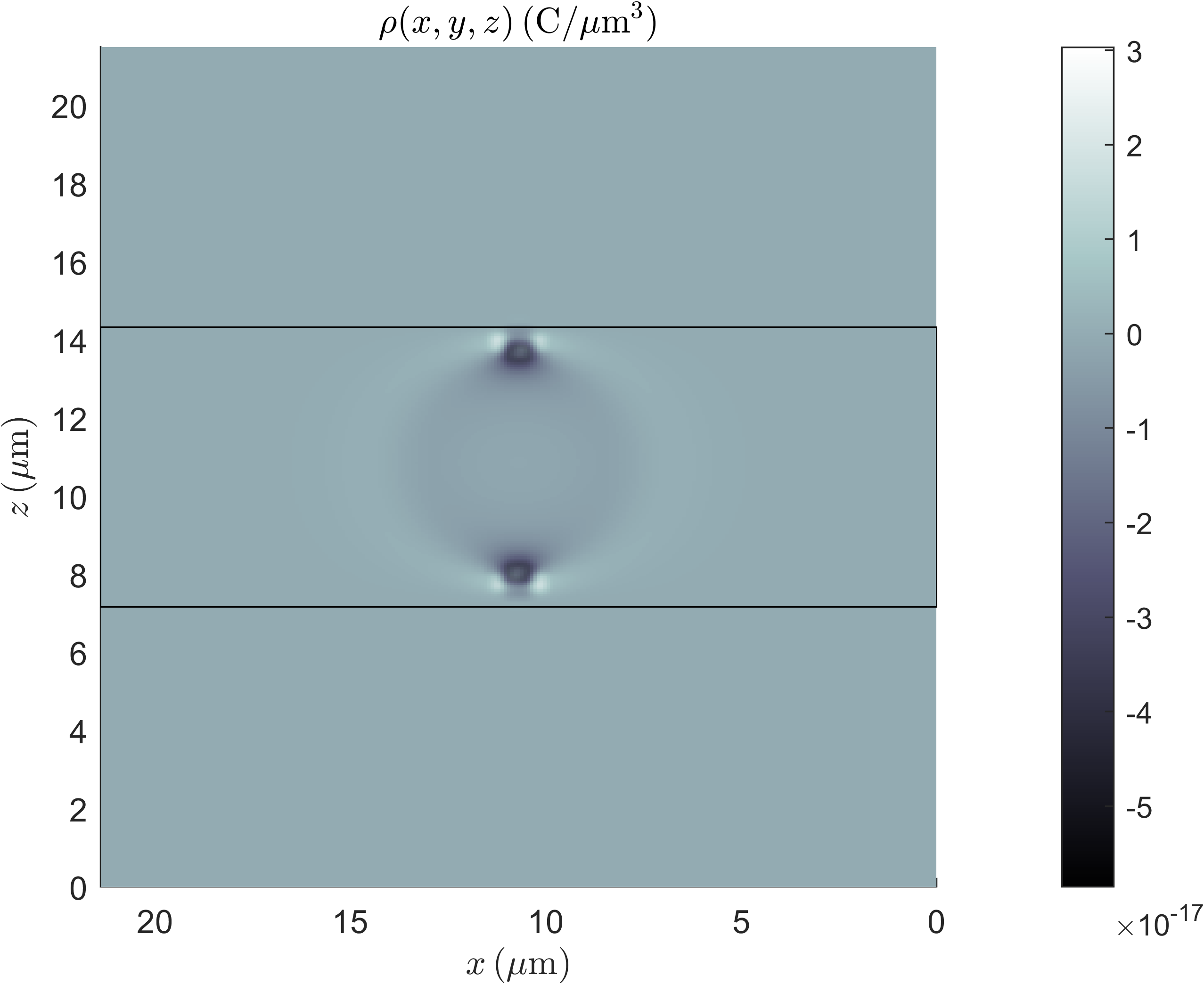}
    \end{subfigure}
    \caption{Transition from a hopfion to a toron due to the flexoelectric self-polarization effect in a liquid crystal in the one constant approximation. The top row shows the self-induced electrostatic scalar potential $\varphi$, which is a solution of the Poisson equation $\Delta\varphi=\rho/\epsilon_0$, and the bottom row displays the electric charge density $\rho=-\bm{\nabla}\cdot\mathbf{P}_f$. It is clear to see that the electric charge density is confined within the confined geometry $\Omega$, whereas the electric scalar potential extends into $\mathbb{R}^3/\Omega$. The parameter set is detailed in Figure \ref{fig: Hopfion results - Director field} with flexoelectric coefficients $e_1=e_3=8\,\textup{pCm}^{-1}$.}
    \label{fig: Skyrmion results - Electric properties}
\end{figure*}

As the flexoelectric coefficient strength is increased from $e_i=4\,\textup{pCm}^{-1}$ to $e_i=8\,\textup{pCm}^{-1}$, the hopfion collapses and transitions to a toron state.
This toron terminates at point defects due to the boundary conditions enforced by the strong homeotropic anchoring \eqref{eq: Strong homeotropic anchoring} \cite{Tai_2018}, and can be seen in Fig.~\ref{fig: Director field - Skyrmion}.
It is clear from Fig.~\ref{fig: Linking skyrmion} that the knotted preimages of $\mathbf{n}=\pm\mathbf{e}_1$ become unlinked.
The resulting electric scalar potential and electric charge density for the toron are plotted in Fig.~\ref{fig: Skyrmion results - Electric properties}.

\section{Conclusion}
\label{sec: Conclusion}

In this paper we have studied the back-reaction of the self-induced flexoelectric polarization on topological defects in chiral liquid crystals.
Due to flexoelectricity, topological defects induce spatially-varying strain into the liquid crystal, which generates an internal polarization \eqref{eq: Flexoelectric polarization}.
From the polarization emerges an electric field \eqref{eq: Electric field} with an associated electrostatic self-energy \eqref{eq: Electrostatic self-energy}.
We show how to compute this self-energy and its corresponding back-reaction on the director field by introducing the electric scalar potential \eqref{eq: Electric scalar potential} and solving the Poisson equation \eqref{eq: Poisson equation}.

The problem we have investigated here is the analogue problem of demagnetization in chiral magnetic systems.
While the two physical systems are similar, the manifestation of topological defects in each system is unique.
Further, the electrostatic self-interaction behaves differently in both systems.
In chiral magnets, the electrostatic scalar potential depends on the divergence of the order parameter, whereas it depends on the divergence of the polarization induced by the order parameter in chiral liquid crystals.
We showed that, in two-dimensional chiral liquid crystal systems, the electrostatic self-energy is conformally invariant.
However, in two-dimensional chiral magnetic systems, the demagnetizing self-energy behaves like a potential and can provide scale stability to magnetic skyrmions and aid in evading the Hobart--Derrick Theorem.
Finally, magnetic Bloch skyrmions are unaffected by the electrostatic self-interaction due to the Bloch ansatz being solenoidal.
Although the Bloch skyrmion is solenoidal, its self-induced polarization is not.
Therefore, liquid crystal Bloch skyrmions are affected by the electrostatic self-interaction.

In addition, we have derived a liquid crystal model that favors splay and bend deformations, and related it to a chiral magnetic system with a DMI term coming from Rashba SOC.
A comparison of the effect of the electrostatic self-interaction on skyrmions in the two liquid crystals models was detailed.
In particular, for unequal flexoelectric coefficients the two skyrmion types yield distinct electric scalar potentials and are not degenerate in energy.
We also showed that increasing the strength of the flexoelectric coefficients can destabilize hopfions into torons in three-dimensional confined systems.


\section*{Acknowledgments}

The author acknowledges funding from the Olle Engkvists Stiftelse through the grant 226-0103 and the Roland Gustafssons Stiftelse för teoretisk fysik.


\appendix


\section{Depolarization}
\label{sec: Depolarization}

Consider the continuous electric dipole moment distribution $\mathbf{P}_f: \Omega \rightarrow \mathbb{R}^3$, where $\Omega \subseteq \mathbb{R}^3$ is the confined geometry \eqref{eq: Confined geometry}.
The electric potential $\varphi:\mathbb{R}^3\rightarrow\mathbb{R}$ associated to this continuous dipole distribution, at a point $\mathbf{x}\in\mathbb{R}^3$, is given by
\begin{equation}
\label{eq: Electric scalar potential}
    \varphi(\mathbf{x}) = \frac{1}{4\pi\epsilon_0} \int_\Omega \textup{d}^3\mathbf{y} \, \left\{ \frac{\mathbf{P}_f(\mathbf{y}) \cdot(\mathbf{x}-\mathbf{y})}{|\mathbf{x}-\mathbf{y}|^3} \right\},
\end{equation}
where $\epsilon_0$ is the permittivity of free space.
The corresponding induced electric field $\mathbf{E}:\mathbb{R}^3\rightarrow\mathbb{R}^3$ is the gradient of this electric potential, 
\begin{align}
\label{eq: Electric field}
    \mathbf{E}(\mathbf{x}) = \, & - \bm{\nabla}_{\mathbf{x}}\varphi(\mathbf{x}) \nonumber \\
    = \, & - \frac{1}{4\pi\epsilon_0} \int_\Omega \textup{d}^3\mathbf{y} \frac{1}{|\mathbf{x}-\mathbf{y}|^3} \left\{ \mathbf{P}_f(\mathbf{y}) 
    \right. \nonumber \\
    \, & \left.- 3 \frac{\mathbf{P}_f(\mathbf{y}) \cdot (\mathbf{x}-\mathbf{y})}{|\mathbf{x}-\mathbf{y}|^2} (\mathbf{x}-\mathbf{y}) \right\}.
\end{align}
Let us note that the Green's function for the Laplacian $\Delta = -\nabla^2$ on $\mathbb{R}^3$ is
\begin{equation}
    G(\mathbf{x},\mathbf{y}) = \frac{1}{4\pi|\mathbf{x}-\mathbf{y}|}, \quad \Delta_{\mathbf{x}} G(\mathbf{x},\mathbf{y}) = \delta(\mathbf{x}-\mathbf{y}).
\end{equation}
Then, using the identity
\begin{equation}
    \bm{\nabla}_{\mathbf{x}} \left(\frac{1}{|\mathbf{x}-\mathbf{y}|}\right) = -\frac{\mathbf{x}-\mathbf{y}}{|\mathbf{x}-\mathbf{y}|^3},
\end{equation}
we can express the electric potential $\varphi$ in terms of the Green's function
\begin{equation}
\label{eq: Electric potential definition}
    \varphi(\mathbf{x}) = -\frac{1}{\epsilon_0} \int_\Omega \textup{d}^3\mathbf{y} \, \left\{ \mathbf{P}_f(\mathbf{y}) \cdot \bm{\nabla}_{\mathbf{x}} G(\mathbf{x},\mathbf{y}) \right\}.
\end{equation}
Now, noting that the gradient and Laplacian commute on $\mathbb{R}^3$, the Laplacian of the electric potential is
\begin{align}
    \Delta_{\mathbf{x}} \varphi(\mathbf{x}) = \, & -\frac{1}{\epsilon_0} \int_\Omega \textup{d}^3\mathbf{y} \, \mathbf{P}_f(\mathbf{y}) \cdot \bm{\nabla}_{\mathbf{x}} \Delta_{\mathbf{x}}G(\mathbf{x},\mathbf{y}) \nonumber \\
    = \, & -\frac{1}{\epsilon_0} \int_\Omega \textup{d}^3\mathbf{y} \, \mathbf{P}_f(\mathbf{y}) \cdot \bm{\nabla}_{\mathbf{x}} \delta(\mathbf{x}-\mathbf{y}) \nonumber \\
    = \, & \frac{1}{\epsilon_0} \int_\Omega \textup{d}^3\mathbf{y} \, \mathbf{P}_f(\mathbf{y}) \cdot \bm{\nabla}_{\mathbf{y}} \delta(\mathbf{x}-\mathbf{y}) \nonumber \\
    = \, & -\frac{1}{\epsilon_0} \int_\Omega \textup{d}^3\mathbf{y} \, \bm{\nabla}_{\mathbf{y}} \cdot \mathbf{P}_f(\mathbf{y}) \delta(\mathbf{x}-\mathbf{y}) \nonumber \\
    = \, & -\frac{1}{\epsilon_0} \bm{\nabla}_{\mathbf{x}} \cdot \mathbf{P}_f(\mathbf{x}).
\end{align}
Therefore, we see that the electric potential $\varphi$ satisfies Poisson's equation for electrostatics \cite{Zavvou_2022,Yang_2022}
\begin{equation}
    \Delta\varphi = -\nabla^2\varphi = -\frac{1}{\epsilon_0} \bm{\nabla}\cdot\mathbf{P}_f.
\end{equation}


\section{Evading the Hobart--Derrick theorem}
\label{sec: Derrick scaling}


\subsection{Rescaling the flexoelectric energy}
\label{subsec: Rescaling the flexoelectric energy}

The flexoelectric energy is intrinsically non-local.
However, we have shown how to overcome this problem by introducing a scalar electric potential, associated to the self-induced polarization.
Now, we want to understand how the flexoelectric self-energy transforms under a coordinate rescaling, to see the role it plays in evading the Hobart--Derrick theorem.
Recall that the energy of a chiral liquid crystal including the electrostatic self-energy is given by the flexoelectric Frank--Oseen (FFO) energy,
\begin{align}
\label{eq: Full static energy}
    F_{\textup{FFO}} = \, & F_{\textup{FO}} + F_{\textup{flexo}} \nonumber \\
    = \, &  \int_{\Omega} \textup{d}^3x \left\{ \frac{K_{1}}{2} (\bm{\nabla}\cdot \mathbf{n})^2 +  \frac{K_{2}}{2} \left[ \mathbf{n} \cdot (\bm{\nabla}\times\mathbf{n}) \right]^2 \right. \nonumber \\
    \, & \left. + \frac{K_{3}}{2} (\mathbf{n} \times \bm{\nabla} \times \mathbf{n})^2 + K_2 q_0\left[ \mathbf{n} \cdot (\bm{\nabla}\times\mathbf{n}) \right] \right. \nonumber \\
    \, & \left. + V(\mathbf{n}) + \frac{1}{2}\mathbf{P}_f \cdot \bm{\nabla}\varphi \right\},
\end{align}
subject to the constraint
\begin{equation}
\label{eq: Poisson equation constraint}
    \Delta \varphi = -\frac{1}{\epsilon_0} \bm{\nabla} \cdot \mathbf{P}_f, \quad \mathbf{P}_f = e_1 \left[ (\bm{\nabla} \cdot \mathbf{n}) \mathbf{n} \right] + e_3 (\mathbf{n} \times \bm{\nabla} \times \mathbf{n}).
\end{equation}
The numerical method developed to study the effect of dipole-dipole interactions on skyrmions in chiral magnets can be used to address this constrained minimization problem \cite{Leask_Speight_2025}.

First of all, we choose to renormalize energy and length scales such that the energy is dimensionless.
This will also make numerical simulations more palatable.
Let us consider an energy and length rescaling with $E=E_0\hat{E}$ and $x=L_0\hat{x}$.
We choose to set our length and energy scales as
\begin{equation}
    L_0 = \frac{1}{q_0}\frac{K_1}{K_2}, \quad E_0 = \frac{1}{q_0}\frac{K_1^2}{K_2}.
\end{equation}
Then the rescaled energy is 
\begin{align}
    \hat{F}_{\textup{FFO}} = \, & \int_{\Omega} \textup{d}^3x \left\{ \frac{1}{2} (\bm{\nabla}\cdot \mathbf{n})^2 + \frac{1}{2} \frac{K_2}{K_1} \left[ \mathbf{n} \cdot (\bm{\nabla}\times\mathbf{n}) \right]^2 \right. \nonumber \\
        \, & \left. + \frac{1}{2} \frac{K_3}{K_1} (\mathbf{n} \times \bm{\nabla} \times \mathbf{n})^2 + \left[ \mathbf{n} \cdot (\bm{\nabla}\times\mathbf{n}) \right] \right.\nonumber \\
        \, & \left. + \frac{1}{q_0^2} \frac{K_1}{K_2^2} V(\mathbf{n})\right\} + \hat{F}_{\textup{flexo}},
\end{align}
where we still need to determine the dimensionless flexoelectric energy $\hat{F}_{\textup{flexo}}$.

Let us consider the rescaling of the electric potential $\varphi = \lambda \hat\varphi$, where $\lambda$ has the units $[\lambda]=\textup{Nm}\textup{C}^{-1}$.
We find that the rescaled flexoelectric energy is
\begin{align}
    \hat{F}_{\textup{flexo}} = \, & \frac{1\epsilon_0}{2E_0} \int_{\Omega} \lambda^2 \hat\varphi \frac{1}{L_0^2} \Delta_{\hat{x}} \hat\varphi L_0^3 \, \textup{d}^3\hat{x} \nonumber \\
    = \, & \frac{1}{2}\frac{L_0\lambda^2\epsilon_0}{E_0} \int_{\Omega} \hat\varphi \Delta_{\hat{x}} \hat\varphi \, \textup{d}^3\hat{x}.
\end{align}
It will be convenient to express the polarization $\mathbf{P}_f$ in terms of a rescaled polarization $\mathbf{P}$, where
\begin{equation}
    \mathbf{P}_f = \frac{e_1}{L_0} \mathbf{P}, \quad \mathbf{P} =  (\bm{\nabla}_{\hat{x}} \cdot \mathbf{n}) \mathbf{n}  + \frac{e_3}{e_1} [\mathbf{n} \times (\bm{\nabla}_{\hat{x}} \times \mathbf{n})].
\end{equation}
Then the rescaled Poisson equation can be given in the form
\begin{equation}
    \Delta_{\hat{x}} \hat\varphi = -\frac{e_1}{\epsilon_0\lambda} \bm{\nabla}_{\hat{x}} \cdot \mathbf{P}.
\end{equation}
Furthermore, let us introduce the dimensionless vacuum electric permittivity
\begin{equation}
    \epsilon = \frac{L_0\lambda^2\epsilon_0}{E_0} = \left(\frac{e_1}{\epsilon_0\lambda}\right)^{-1}.
\end{equation}
It can be seen that the necessary electric potential rescaling is given by $\lambda=K_1/e_1$ and, hence, the dimensionless vacuum electric permittivity is found to be
\begin{equation}
    \epsilon = \frac{K_1\epsilon_0}{e_1^2}.
\end{equation}
Therefore, the rescaled flexoelectric energy is determined to be
\begin{equation}
    \hat{F}_{\textup{flexo}} = \frac{\epsilon}{2} \int_{\Omega} \hat\varphi \Delta_{\hat{x}} \hat\varphi \, \textup{d}^3\hat{x}, \quad \Delta_{\hat{x}} \hat\varphi = -\frac{1}{\epsilon} \bm{\nabla}_{\hat{x}} \cdot \mathbf{P}.
\end{equation}


\subsection{Scale invariance of the flexoelectric energy in two dimensions}
\label{subsec: Derrick scaling}

We now focus on the situation at hand: translation invariant configurations in the $\mathbf{e}_z$ direction.
Recall that we are considering the finite energy of a thick slab $\Omega=\mathbb{R}^2\times[0,1]$.
Under the energy and length rescalings \eqref{eq: Rescaled energy/length}, the thick slab gets mapped to $\Omega'=\mathbb{R}^2\times[0,t]$, where $t=L_0^{-1}$ is the rescaled slab thickness.
We then consider the energy per unit length $F/t$ of this system, which is defined by the two-dimensional adimensional energy functional
\begin{align}
    F_{\textup{FFO}} = \, & \int_{\mathbb{R}^2} \textup{d}^2x \left\{ \frac{1}{2} (\bm{\nabla}\cdot \mathbf{n})^2 + \frac{1}{2} \frac{K_2}{K_1} \left[ \mathbf{n} \cdot (\bm{\nabla}\times\mathbf{n}) \right]^2 \right. \nonumber \\
    \, & \left. + \frac{1}{2} \frac{K_3}{K_1} (\mathbf{n} \times \bm{\nabla} \times \mathbf{n})^2 + \left[ \mathbf{n} \cdot (\bm{\nabla}\times\mathbf{n}) \right] \right.\nonumber \\
    \, & \left. + \frac{1}{q_0^2} \frac{K_1}{K_2^2} V(\mathbf{n}) + \frac{\epsilon}{2}\varphi\Delta\varphi \right\},
\end{align}
where the electric potential satisfies the dimensionless Poisson equation
\begin{equation}
    \Delta \varphi = -\frac{1}{\epsilon} \bm{\nabla} \cdot \mathbf{P}, \quad \mathbf{P} =  (\bm{\nabla} \cdot \mathbf{n}) \mathbf{n}  + \frac{e_3}{e_1} \left[\mathbf{n} \times (\bm{\nabla} \times \mathbf{n})\right].
\end{equation}

A necessary requirement for topological solitons to exist in a field theory is the successful evasion of the Hobart--Derrick theorem \cite{Derrick_1964}.
The Hobart--Derrick theorem is a non-existence theorem that states if the energy functional $F[\mathbf{n}]$ is not stationary against spatial rescaling, then $\mathbf{n}$ cannot be a solution of the field equations.
So, let us consider a coordinate rescaling $\mathbf{x} \mapsto \mathbf{x}'=\mu\mathbf{x}$, for some positive scaling parameter $\mu\in\mathbb{R}_{>0}$.
Then the director field necessarily rescales as $\mathbf{n}_\mu=\mathbf{n}(\mu\mathbf{x})$.
Obviously, the gradient transforms as $\bm{\nabla} \mapsto \bm{\nabla}'=\frac{1}{\mu}\bm{\nabla}$ under the coordinate rescaling.
To determine how the electric potential $\varphi$ transforms under the coordinate rescaling, we first consider how the polarization $\mathbf{P}$ transforms, which is
\begin{align}
    \mathbf{P}_\mu = \, & \left(\mu\bm{\nabla}' \cdot \mathbf{n}_\mu\right) \mathbf{n}_\mu  + \frac{e_3}{e_1} \left(\mathbf{n}_\mu \times \mu\bm{\nabla}' \times \mathbf{n}_\mu\right) \nonumber \\
    = \, & \mu \left(\bm{\nabla}' \cdot \mathbf{n}(\mu\mathbf{x})\right) \mathbf{n}(\mu\mathbf{x})  + \mu\frac{e_3}{e_1} \left(\mathbf{n}(\mu\mathbf{x}) \times \bm{\nabla}' \times \mathbf{n}(\mu\mathbf{x})\right) \nonumber \\
    = \, & \mu \mathbf{P}(\mu\mathbf{x}).
\end{align}
Now, consider the Poisson equation \eqref{eq: Adimensional 2D Poisson equation}, which has the scaling behavior
\begin{equation}
    \Delta' \varphi_\mu = -\frac{1}{\mu\epsilon} \bm{\nabla}' \cdot \mathbf{P}_\mu = -\frac{1}{\epsilon} \bm{\nabla}' \cdot \mathbf{P}(\mu\mathbf{x}) = \Delta'\varphi(\mu\mathbf{x}).
\end{equation}
It is clear to see that the electric scalar potential scaling behavior is $\varphi_\mu(\mathbf{x})=\varphi(\mu\mathbf{x})$.
Therefore, under the coordinate rescaling $\mathbf{x} \mapsto \mathbf{x}'=\mu\mathbf{x}$, the dimensionless energy functional becomes
\begin{align}
    F_{\textup{FFO}}(\mu) = \, & \int_{\mathbb{R}^2} \textup{d}^2x' \left\{ \frac{1}{2} (\bm{\nabla}'\cdot \mathbf{n})^2 + \frac{1}{2} \frac{K_2}{K_1} \left[ \mathbf{n} \cdot (\bm{\nabla}'\times\mathbf{n}) \right]^2 \right. \nonumber \\
    \, & \left. + \frac{1}{2} \frac{K_3}{K_1} (\mathbf{n} \times \bm{\nabla}' \times \mathbf{n})^2 + \frac{1}{\mu}\left[ \mathbf{n} \cdot (\bm{\nabla}'\times\mathbf{n}) \right] \right.\nonumber \\
    \, & \left. + \frac{1}{\mu^2}\frac{1}{q_0^2} \frac{K_1}{K_2^2} V(\mathbf{n}) + \frac{1}{2}\mathbf{P}\cdot\bm{\nabla}'\varphi \right\} \nonumber \\
    = \, & F_2 + \frac{1}{\mu}F_1 + \frac{1}{\mu^2} F_0 + F_{\textup{flexo}}.
\end{align}
For skyrmions to exist in this model, we require the energy to be stable against spatial rescalings, which yields the Derrick scaling constraint
\begin{equation}
    \left.\frac{\textup{d}F_{\textup{FFO}}}{\textup{d}\mu}\right|_{\mu=1} = -(F_1 + 2F_0) = 0.
\end{equation}
While the potential energy $F_0$ is positive semi-definite, the first order (in spatial derivatives) term $F_1$ can be negative, and thus provide stability.
Hence, the Hobart--Derrick non-existence theorem can be evaded.

We note that the flexoelectric self-energy is scale invariant in two dimensions and is thus unable to provide stability against spatial rescalings.
Whereas, in comparison with chiral ferromagnets, the magnetostatic self-energy there can stabilize skyrmions as it behaves like a potential under coordinate rescalings \cite{Leask_Speight_2025}.


\section{Variation of the flexoelectric energy}

So far, we have shown how to include the electrostatic self-energy and compute the electric scalar potential $\varphi$ by solving Poisson's equation \eqref{eq: Adimensional 2D Poisson equation} for fixed director field configuration $\mathbf{n}$.
Additionally, we have determined that the flexoelectric energy is conformally invariant in two dimensions.
However, we need to compute the back-reaction of the self-induced electric field $\mathbf{E}$ on the director field $\mathbf{n}$.
To do this, we need to calculate the first variation of the flexoelectric energy $F_{\textup{flexo}}(\mathbf{n})$ with respect to the director field $\mathbf{n}$.


\subsection{Two-dimensional system}

Let $\mathbf{n}_t$ be a smooth variation of $\mathbf{n}=\mathbf{n}_0$ through fields of compact support and define
$\delta\mathbf{n}=\partial_t\mathbf{n}_t|_{t=0}$.
Denote by $\varphi_t$ the associated unique solution of \eqref{eq: Adimensional 2D Poisson equation}
with source $-\frac{1}{\epsilon} \bm{\nabla}\cdot \mathbf{P}_t$ decaying to $0$ at infinity, and $\dot\varphi=\partial_t\varphi_t|_{t=0}$.
It is important to note that, while $\delta\mathbf{n}$ has compact support, neither $\varphi=\varphi_0$ nor $\dot\varphi$ do: as argued above, they are $1/r$ localized.
Following \cite{Leask_Speight_2025}, the variation of $F_{\textup{flexo}}$ induced by $\mathbf{n}_t$ is
\begin{align}
    \frac{\textup{d}}{\textup{d}t}\bigg|_{t=0}F_{\textup{flexo}}(\mathbf{n}_t) = \, & \frac{\epsilon}{2} \int_{\mathbb{R}^2} \textup{d}^2x \left(\dot\varphi\Delta\varphi+\varphi\Delta\dot\varphi\right).
\end{align}
Let us consider the term
\begin{align}
    \int_{\mathbb{R}^2} \textup{d}^2x \, \dot\varphi\Delta\varphi = \, & - \int_{\mathbb{R}^2} \textup{d}^2x \, \dot\varphi (\bm{\nabla} \cdot \bm{\nabla}\varphi) \nonumber \\
    = \, & \int_{\mathbb{R}^2} \textup{d}^2x \, \braket{\bm{\nabla}\varphi, \bm{\nabla}\dot\varphi} - \int_{\mathbb{R}^2} \textup{d}^2x \, \bm{\nabla} \cdot \left( \dot\varphi \bm{\nabla}\varphi \right) \nonumber \\
    = \, & \int_{\mathbb{R}^2} \textup{d}^2x \, \varphi\Delta\dot\varphi + \int_{\mathbb{R}^2} \textup{d}^2x \, \bm{\nabla} \cdot \left( \varphi \bm{\nabla}\dot\varphi - \dot\varphi \bm{\nabla}\varphi \right) \nonumber \\
    = \, & \int_{\mathbb{R}^2} \textup{d}^2x \, \varphi\Delta\dot\varphi \nonumber \\
    \, & - \oint_{\partial B_{\infty}(0)} \textup{d}\mathbf{s} \cdot \left( \varphi \bm{\nabla}\dot\varphi - \dot\varphi \bm{\nabla}\varphi \right).
 \end{align}
So, the variation of the electrostatic self-energy is found to be given by
\begin{align}
    \frac{\textup{d}}{\textup{d}t}\bigg|_{t=0}F_{\textup{flexo}}(\mathbf{n}_t) = \, & \epsilon\int_{\Omega}\textup{d}^3x\, \varphi\Delta\dot\varphi \nonumber \\
    \, & - \frac{\epsilon}{2}\oint_{\partial B_{\infty}(0)} \textup{d}\mathbf{s} \cdot \left( \varphi \bm{\nabla}\dot\varphi - \dot\varphi \bm{\nabla}\varphi \right).
\end{align}
The surface term vanishes
\begin{align}
    \dot{F}_{\textup{flexo}}^{\textup{surf}} = \, & -\frac{\epsilon}{2}\oint_{\partial B_{\infty}(0)} \textup{d}\mathbf{s} \cdot \left( \varphi \bm{\nabla}\dot\varphi - \dot\varphi \bm{\nabla}\varphi \right) \nonumber \\
    = \, & \lim_{R\rightarrow\infty}\frac{\epsilon}{2}\int_{\partial B_R(0)}(\dot\varphi\star\textup{d}\varphi-\varphi \star\textup{d}\dot\varphi) \nonumber \\
    = \, & \lim_{R\rightarrow\infty}\frac{\epsilon R}{2}\int_0^{2\pi}(\dot\varphi\varphi_r-\varphi\dot\varphi_r) \textup{d}\theta \nonumber \\
    = \, & 0,
\end{align}
since $\varphi,\dot\varphi=O(r^{-1})$ and $\varphi_r,\dot\varphi_r=O(r^{-2})$.
Varying the Poisson equation \eqref{eq: Adimensional 2D Poisson equation} with respect to $t$, we deduce that
\begin{equation}
    \Delta\dot\varphi = -\frac{1}{\epsilon} \bm{\nabla}\cdot\left(\frac{\textup{d}}{\textup{d}t}\bigg|_{t=0}\mathbf{P}(\mathbf{n}_t)\right),
\end{equation}
where the variation in the flexoelectric polarization is
\begin{align}
    \frac{\textup{d}}{\textup{d}t}\bigg|_{t=0}\mathbf{P}(\mathbf{n}_t) 
    = \, & \frac{e_3}{e_1} \left[\delta\mathbf{n} \times (\bm{\nabla} \times \mathbf{n}) + \mathbf{n} \times (\bm{\nabla} \times \delta\mathbf{n}) \right] \nonumber \\
    \, & +(\bm{\nabla} \cdot \delta\mathbf{n}) \mathbf{n} + (\bm{\nabla} \cdot \mathbf{n}) \delta\mathbf{n}.
\end{align}
Then, the variation of the flexoelectric energy is
\begin{align}
    \frac{\textup{d}}{\textup{d}t}\bigg|_{t=0}F_{\textup{flexo}}(\mathbf{n}_t) = \, & -\int_{\mathbb{R}^2}\textup{d}^2x\, \varphi\bm{\nabla}\cdot\left(\frac{\textup{d}}{\textup{d}t}\bigg|_{t=0}\mathbf{P}(\mathbf{n}_t)\right) \nonumber \\
    = \, & \int_{\mathbb{R}^2}\textup{d}^2x\, \bm{\nabla}\varphi \cdot \left(\frac{\textup{d}}{\textup{d}t}\bigg|_{t=0}\mathbf{P}(\mathbf{n}_t)\right) \nonumber \\
    = \, & \int_{\mathbb{R}^2}\textup{d}^2x\, (\textup{grad}_{\mathbf{n}}\,F_{\textup{flexo}}) \cdot \delta\mathbf{n},
\end{align}
where the corresponding gradient is
\begin{align}
    \textup{grad}_{\mathbf{n}}\,F_{\textup{flexo}} = \, & \frac{e_3}{e_1} \left[  \left( (\bm{\nabla} \times \mathbf{n}) \times \bm{\nabla}\varphi \right) + \left( \bm{\nabla} \times (\bm{\nabla}\varphi\times\mathbf{n}) \right) \right] \nonumber \\
    \, & - \bm{\nabla}(\bm{\nabla}\varphi \cdot \mathbf{n}) + (\bm{\nabla}\cdot\mathbf{n})\bm{\nabla}\varphi.
\end{align}
To arrive at this, we have used the divergence theorem and the vector calculus identities
\begin{equation}
    (\bm{\nabla}\varphi\cdot\mathbf{n})(\bm{\nabla}\cdot\delta\mathbf{n}) = \bm{\nabla} \cdot \left[ (\bm{\nabla}\varphi \cdot \mathbf{n})\delta\mathbf{n} \right] - \bm{\nabla}(\bm{\nabla}\varphi \cdot \mathbf{n}) \cdot \delta\mathbf{n}
\end{equation}
and
\begin{align}
    (\bm{\nabla}\varphi\times\mathbf{n})\cdot(\bm{\nabla}\times\delta\mathbf{n}) = \, & \left[ \bm{\nabla} \times (\bm{\nabla}\varphi\times\mathbf{n}) \right] \cdot \delta\mathbf{n} \nonumber \\
    \, & - \bm{\nabla} \cdot \left[ (\bm{\nabla}\varphi\times\mathbf{n})\times\delta\mathbf{n} \right].
\end{align}


\bibliography{main.bib}

\end{document}